\documentclass[fleqn,usenatbib]{mnras}
\usepackage{newtxtext,newtxmath}

\usepackage[T1]{fontenc}
\usepackage{ae,aecompl}
\usepackage{times}
\usepackage{rotating}
\usepackage{graphicx}	
\usepackage{amsmath,amstext,bm}	
\usepackage[usenames]{xcolor}

\usepackage{soul}

\newcommand{\Msun}{\,{\rm M_{\odot}}}

\newcommand{\epsff}{\epsilon_{\mathrm{ff}}}

\newcommand{\tff}{\tau_{\mathrm{ff}}}

\newcommand{\mmax}{M_{\mathrm{max}}}
\newcommand{\mmin}{M_{\mathrm{min}}}

\newcommand{\ltid}{\lambda_{\mathrm{tid}}}
\newcommand{\ltidfifty}{\lambda_{\mathrm{tid, 50}}}
\newcommand{\ttid}{t_{\mathrm{tid}}}
\newcommand{\Ssfr}{\Sigma_{\mathrm{SFR}}}
\newcommand{\rgal}{R_{\mathrm{gal}}}

\title[Young massive clusters in galaxy mergers]{Formation and evolution of young massive clusters in galaxy mergers: the \texttt{SMUGGLE} view}

\author[Hui~Li et al.]
    {\parbox[T]{18cm}{Hui Li$^{1,2}$\thanks{E-mail: li.hui@columbia.edu}\thanks{NHFP Hubble Fellow}\href{http://orcid.org/0000-0002-1253-2763}{\includegraphics[scale=0.8]{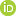}},
    Mark Vogelsberger$^{2}$\href{http://orcid.org/0000-0001-8593-7692}{\includegraphics[scale=0.8]{figure/orcid.png}},
    Greg L. Bryan$^{1,3}$\href{http://orcid.org/0000-0003-2630-9228}{\includegraphics[scale=0.8]{figure/orcid.png}},
    Federico Marinacci$^{4}$\href{http://orcid.org/0000-0003-3816-7028}{\includegraphics[scale=0.8]{figure/orcid.png}},
    Laura V. Sales$^{5}$\href{https://orcid.org/0000-0002-3790-720X}{\includegraphics[scale=0.8]{figure/orcid.png}},
    Paul Torrey$^{6}$\href{http://orcid.org/0000-0002-5653-0786}{\includegraphics[scale=0.8]{figure/orcid.png}}
     }\vspace{0.2cm}\\
     $^{1}$Department of Astronomy, Columbia University, New York, NY 10027, USA\\
     $^{2}$Department of Physics and Kavli Institute for Astrophysics and Space Research, Massachusetts Institute of Technology, Cambridge, MA 02139, USA \\
     $^{3}$Center for Computational Astrophysics, Flatiron Institute, 162 5th Ave, New York, NY 10003, USA\\
     $^{4}$Department of Physics \& Astronomy ``Augusto Righi'', University of Bologna, via Gobetti 93/2, 40129 Bologna, Italy\\
     $^{5}$Department of Physics \& Astronomy, University of California, Riverside, 900 University Avenue, Riverside, CA 92521, USA\\
     $^{6}$Department of Astronomy, University of Florida, 211 Bryant Space Sciences Center, Gainesville, FL 32611, USA\\
    }
    
\date{Accepted XXX. Received YYY; in original form ZZZ}

\pubyear{2022}

\begin{document}
\label{firstpage}
\pagerange{\pageref{firstpage}--\pageref{lastpage}}
\maketitle

\begin{abstract}
Galaxy mergers are known to host abundant young massive cluster (YMC) populations, whose formation mechanism is still not well-understood. Here, we present a high-resolution galaxy merger simulation with explicit star formation and stellar feedback prescriptions to investigate how mergers affect the properties of the interstellar medium and YMCs. Compared with a controlled simulation of an isolated galaxy, the mass fraction of dense and high-pressure gas is much higher in mergers. Consequently, the mass function of both molecular clouds and YMCs becomes shallower and extends to higher masses. Moreover, cluster formation efficiency is significantly enhanced and correlates positively with the star formation rate surface density and gas pressure. We track the orbits of YMCs and investigate the time evolution of tidal fields during the course of the merger. At an early stage of the merger, the tidal field strength correlates positively with YMC mass, $\ltid\propto M^{0.71}$, which systematically affects the shape of the mass function and age distribution of the YMCs. At later times, most YMCs closely follow the orbits of their host galaxies, gradually sinking into the center of the merger remnant due to dynamical friction, and are quickly dissolved via efficient tidal disruption. Interestingly, YMCs formed during the first passage, mostly in tidal tails and bridges, are distributed over a wide range of galactocentric radii, greatly increasing their survivability because of the much weaker tidal field in the outskirts of the merger system. These YMCs are promising candidates for globular clusters that survive to the present day.
\end{abstract}




\section{Introduction}\label{sec:intro}

The interactions of dark matter halos over cosmic time are the key physical process of hierarchical structure formation in the $\Lambda$-cold dark matter cosmology \citep{somerville_dave15}. Mergers of halos naturally lead to the mergers of their central galaxies and inevitably leave strong imprints on their evolution.

Over the last several decades, observations of nearby and high redshift galaxies suggest that major mergers alter star formation activities \citep[e.g.][]{ellison_etal08, wong_etal11, patton_etal13, knapen_etal15, davies_etal15}, modify the morphology of galaxies \citep[e.g.][]{conselice_etal03, lotz_etal08, mortlock_etal13}, trigger the formation of ultraluminous infrared galaxies \citep[e.g.][]{sanders_etal88,melnick_mirabel90}, and enhance atomic/molecular gas contents \citep[e.g.][]{braine_combes93, ueda_etal14, larson_etal16, violino_etal18, ellison_etal18}. Moreover, major mergers not only change the global properties of the galaxies but also strongly affect various spatially-resolved properties of the interstellar medium (ISM) and modes of star formation. Several high-resolution optical observations of nearby merger systems \citep[e.g.][]{holtzman_etal92, whitmore_schweizer95, holtzman_etal96, schweizer_seitzer98, zepf_etal99, forbes_hau00, schweizer04, reines_etal08, miah_etal15} have discovered a large number of young massive clusters (YMCs) with masses larger than $10^4\Msun$ and ages younger than 100 Myr \citep[see a recent review by][]{portegies_etal10}. The total number of YMCs are much larger and the cluster initial mass function (CIMF), best described as a power-law shape \citep[e.g.][]{zhang_fall99, bik_etal03, bastian_08}, extends to much higher masses in merger systems compared to those in normal disk galaxies.
The most recent HiPEEC survey \citep[][]{adamo_etal20} quantifies the cluster formation process of four interacting galaxies and found that these galaxies have among the highest cluster formation efficiencies (CFE), 
demonstrating that mergers are important for our understanding of the formation mechanisms of YMCs in extreme galactic environments. 

On the theoretical side, investigating the nature and consequences of galaxy mergers dates back to the mid-20th century \citep[e.g.][]{holmberg41, zwicky56, pfleiderer_siedentopf61}.
The seminal work of \citet{toomre_toomre72} quantitatively modeled various galaxy merger configurations and reproduced many key features, such as tidal tails and bridges, that are consistent with observations. \citet[][]{toomre77} later recognized that mergers can strongly influence the kinematics of galaxies and possibly transform galactic disks into objects that resemble elliptical galaxies, an idea that was examined extensively by \citet{negroponte_white83, barnes88,barnes92,hernquist92,hernquist93} in dissipationless simulations.
Subsequent hydrodynamical simulations demonstrated that major mergers can trigger starbursts \citep[e.g.][]{mihos_hernquist94,mihos_hernquist96, barnes04, springel_etal05b, hayward_etal14, di-matteo_etal07, saitoh_etal09}, although the enhancement depends on the types of galaxy \citep[e.g.][]{di-matteo_etal07, cox_etal08}, disk orientation \citep[e.g.][]{naab_burkert03,cox_etal06}, subgrid galaxy formation model \citep[e.g.][]{cox_etal06, robertson_etal06,hopkins_etal13b}, and numerical resolution \citep[e.g.][]{teyssier_etal10}.
More interestingly, recent high resolution hydrodynamic simulations start to resolve the ISM structure and star formation activities in interacting galaxies.
For example, \citet[][]{bournaud_etal11} found that, although radial gas inflow efficiently fuels the nuclear starbursts during the final coalescence of mergers, tidal compression, turbulence, and gas fragmentation that are driven by mergers can trigger more extended starbursts \citep[see also][]{renaud_etal15}. \citet[][]{moreno_etal19,moreno_etal21} quantified the spatial and temporal distribution of the enhancement of star formation using the FIRE2 model and found that mergers not only enhance the formation of cold dense gas, but also the local star formation efficiency.

On the other hand, modeling the formation of giant molecular clouds (GMCs) and YMCs in mergers is still computationally challenging, because resolving the dynamical states of these parsec-scale structures \citep[e.g.][]{oconnell_etal94, mclaughlin_vandermarel05, bastian_etal13, portegies_zwart_etal10} requires extremely high spatial and mass resolution as well as reliable star formation and stellar feedback models.
Encouragingly, after decades of efforts \citep[e.g.][]{bekki_couch01, bekki_chiba02, li_etal04, kravtsov_gnedin05}, several recent theoretical models \citep[e.g.][]{kruijssen_etal12} and idealized galaxy mergers simulations \citep[e.g.][]{bournaud_etal08,matsui_etal12, powell_etal13, renaud_etal15, lahen_etal20} are able to model the statistical properties of the YMCs over a large mass range. These simulations are able to reproduce the formation of most massive YMCs with mass as high as $\sim10^7\Msun$ with a power-law shape CIMF, although controversy still exists \citep[see e.g.][]{renaud_etal15,maji_etal17}.
Most recently, this problem has been examined in a few cosmological simulations that take into account realistic mass assembly histories of the host galaxies under the framework of hierarchical structure formation \citep[e.g.][]{li_etal17,li_etal18, pfeffer_etal18,kim_etal18_fire,ma_etal20,keller_etal20}. Although mainly focusing on high-z galaxies due to unprecedented computing costs, these studies revealed the prevalence of YMCs formed during frequent major mergers at high-$z$ and provide solid evidence that these YMCs are promising candidates for globular cluster progenitors.

Following recent advances of both observations and simulations of spatially-resolved ISM and YMC populations, here we present a controlled experiment of galaxy merger simulations with the novel galaxy formation model, \texttt{SMUGGLE} \citep[][]{marinacci_etal19}. We comprehensively investigate and compare the properties of GMCs and YMCs in galaxy mergers and the isolated galaxy counterpart. The goal of this paper is to quantify the enhancement of the CFE during major mergers, explore the relationship between cluster formation and different galactic environments, and provide insights in to the long-term dynamical disruption process of YMCs. 

The paper is organized as follows. In \autoref{sec:methods}, we briefly summarize the physical processes in the \texttt{SMUGGLE} model, describe the initial conditions of the galaxy merger simulations, and illustrate the workflow to identify GMCs and YMCs from the simulation snapshots. In \autoref{sec:results}, we compare the properties of the star formation and cluster formation activities between isolated and merger runs and quantify the enhancement of the CFE in different galactic environments. Moreover, in \autoref{sec:results-tracking}, we study the orbital and tidal properties of YMCs formed at different evolution stages of the mergers and investigate how mergers affect their survival probabilities. In \autoref{sec:discussion}, we compare our results with some previous studies and derive analytical models of the time evolution of YMC mass function in mass-dependent tidal disruption. Finally, we summarized the key results of the paper in \autoref{sec:summary}. 

\section{Methods}\label{sec:methods}

\begin{figure*}
\includegraphics[width=\textwidth]{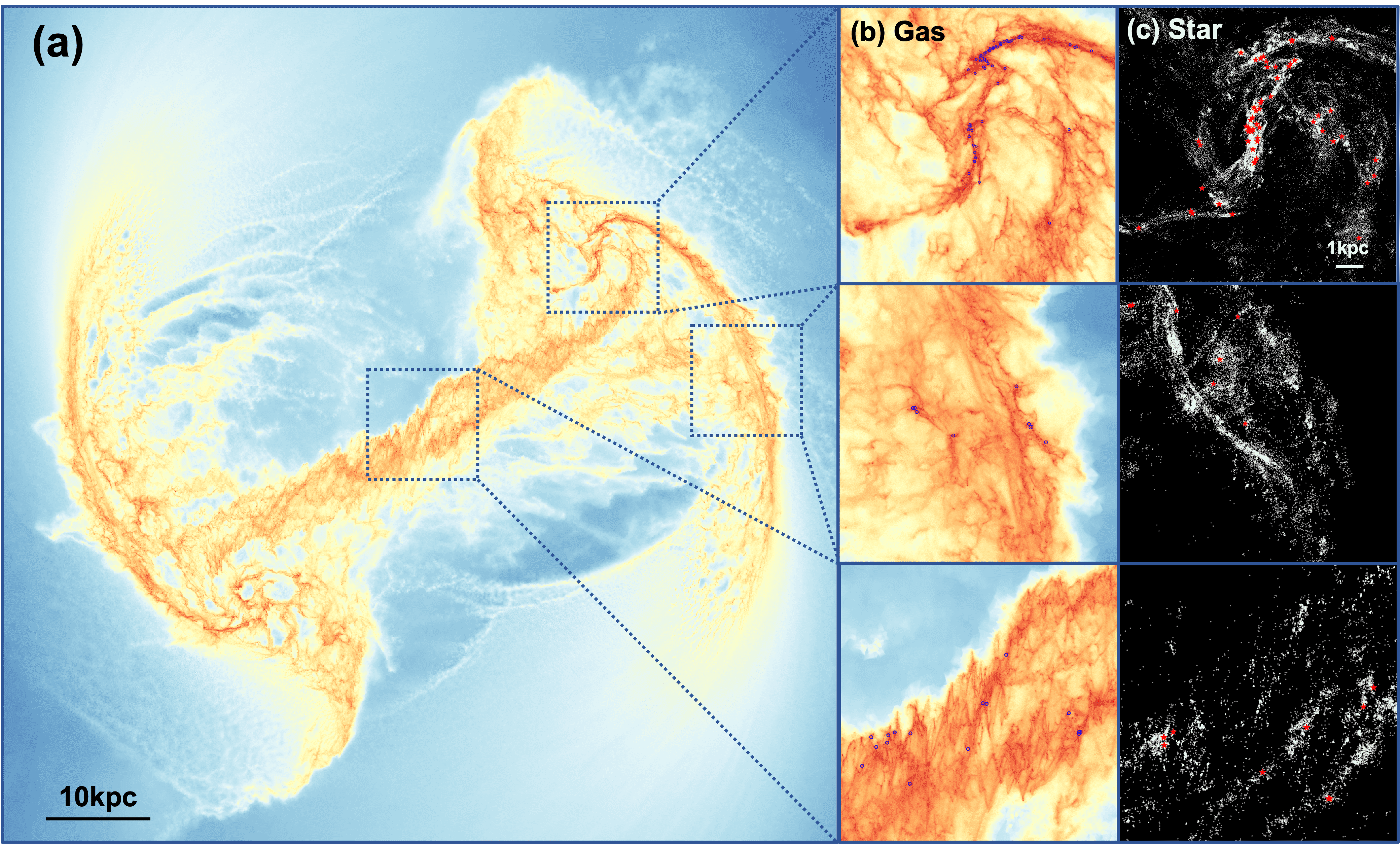}
\vspace{0mm}
\caption{Bound GMCs and YMCs identified in the simulated merger of two Milky Way-sized galaxies after the first passage at $t=0.25$~Gyr. Panel (a): Gas surface density of the merging pair. Strong tidal features, such as tidal tails and bridges, are clearly present. Panel (b): Zoom-in view of the gas distribution in the central region of the host galaxy (upper), in the tidal tail (middle), and in the gas bridge (lower). The identified GMC candidates more massive than $5\times10^5\Msun$ are labeled as blue circles.  Panel (c): the same three zoom-in regions as panel (b) but for the distribution of stars. The identified YMCs more massive than $10^5\Msun$ are labeled as red stars.}
  \label{fig:merger-gas-star}
\end{figure*}

The simulations in this work are performed with \textsc{arepo} \citep{springel10}, a moving-mesh, finite-volume hydrodynamic code with second-order Godunov scheme. Closely following \citet[][]{li_etal20} (L20), the idealized galaxy merger simulations utilize the the Stars and MUltiphase Gas in GaLaxiEs (\texttt{SMUGGLE}) framework \citep{marinacci_etal19}, which includes hydrodynamics, self-gravity, radiative heating/cooling, star formation, and stellar feedback.
To better quantify the effect of the galaxy merger on the star and cluster formation, we compare the results from the merger simulation with a ``control sample'' of an isolated galaxy simulation used in L20.

\subsection{The \texttt{SMUGGLE} galaxy formation model}\label{sec:methods-SMUGGLE}

In order to isolate the impact of major mergers, we keep the model ingredients and parameters the same as the SFE1 run in L20. Here we briefly recap some key physical ingredients in \texttt{SMUGGLE} and the model parameters we used. The model incorporates explicit gas cooling and heating over a large range of temperatures between 10 and $10^8$~K so that the thermodynamic properties of the multi-phase ISM are modelled explicitly. Star particles are formed from cold, dense, and self-gravitating molecular gas at a rate, $\dot{M}_*=\epsff M_{\rm gas}/\tff$, that depends on the mass of the gas cell ($M_{\rm gas}$), the local free-fall time ($\tff$), and the star formation efficiency per free-fall time ($\epsff$). We use a fiducial value 0.01 for the efficiency $\epsff$, the same as the SFE1 run in L20. All relevant stellar feedback processes, such as photoionization, radiation pressure, energy and momentum injection from stellar winds and supernovae (SNe), are included with a Chabrier \citep[][]{chabrier03} initial mass function. The technical details of the \texttt{SMUGGLE} model are extensively described in \citet{marinacci_etal19} and L20. We highlight that the \texttt{SMUGGLE} model successfully reproduces the multiphase ISM structure, generates galactic fountain flows self-consistently, and maintains feedback-regulated inefficient star formation that is consistent with observations.

\subsection{Initial conditions}\label{sec:methods-IC}

In L20, we studied the physical properties of the GMCs in an isolated galaxy, which consisted of a stellar bulge and disc, a gaseous disc, and a dark matter halo, with masses similar to the Milky Way. The physical parameters of the initial conditions in this isolated disk are described in L20. In the current work, we construct new initial conditions of the galaxy merger by putting two identical isolated galaxies together in a pre-merger state. We specify the spin axis of each disc with parameters $\theta$ and $\phi$ in spherical coordinates for each galaxy. We follow the work done by \citet{cox_etal06, hopkins_etal13b} and choose to simulate a near-prograde merger with the disc spin parameters $(\theta_1, \phi_1, \theta_2,\phi_2)=(30^\circ,60^\circ,-30^\circ,45^\circ)$. This type of merger represents a strong resonant interaction so that the resulting tidal features are strongly developed.
Note that the goal of this experiment is to explore the effects of mergers on the formation of GMCs and YMCs, not to fully explore all merger configurations. Therefore, here we only consider equal-mass major mergers with a parabolic orbit \citep[e.g.][]{benson05, khochfar_burkert06}. Systematic studies of different merger configuration can be found in \citet[][]{naab_burkert03, cox_etal06,robertson_etal06, cox_etal08, burkert_etal08, jesseit_etal09}.

The mass resolution for both gas and stars is $\sim1.4\times10^3\Msun$, which allows us to resolve GMCs and YMCs more massive than $4\times10^4\Msun$ with more than 30 resolution elements. The gravitational softening for gas cells is fully adaptive with a minimum softening of $3.6$~pc. We use this same value for the softening length of the star particles as well.

\subsection{Cloud and cluster identification}\label{sec:methods-clustering}
In L20, we used the \textsc{astrodendro} algorithm \citep{rosolowsky_etal08} to identify GMCs in 2D molecular gas surface density maps of the simulated disk galaxies. That method mimics the observational analysis used for sub-mm observations and is crucial when the goal is to closely compare simulations with observations in order to distinguish different model parameters. In contrast, the goal of this paper is to investigate the \textit{physical} effects of galaxy mergers on the spatially-resolved properties of GMCs and YMCs. Therefore, we are more interested in the actual clouds in 3D rather than the ones that are identified in 2D projections.

\begin{figure*}
\includegraphics[width=\textwidth]{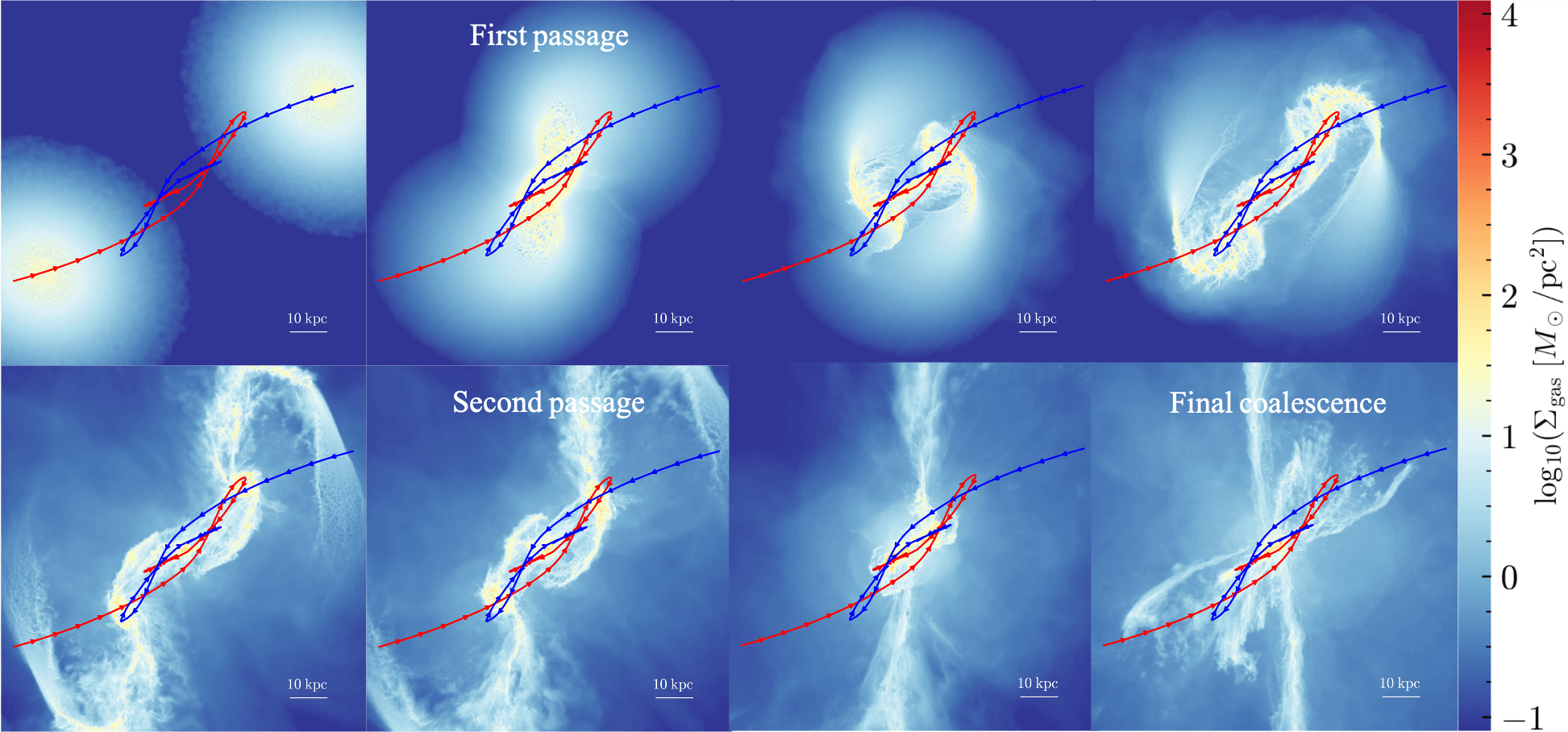}
\vspace{0mm}
\caption{Time series of the simulated merger in different evolutionary stages: e.g. the first passage ($\sim0.22$~Gyr), second passage ($\sim0.52$~Gyr), and final coalescence ($\sim0.69$~Gyr). Red and blue arrows indicate the trajectories of the center-of-mass of the galaxies on the left and right, respectively.}
  \label{fig:merger-series}
\end{figure*}

We identify GMCs from all gas cells via a modified version of SUBFIND \citep[][]{springel_etal01} that take into account adaptive softening lengths when estimating gravitational boundness. We also identify SUBFIND groups of star particles and obtained the \textit{bound} YMC candidates in each snapshot of both the isolated and merger simulations. 
We remove GMCs or YMCs that contains less than 30 resolution elements as these low mass objects suffer low number statistics and strong numerical disruption. The criterion for this cut is somewhat arbitrary, but we find that varying this value by a factor of few only affects the lower-mass end of the mass function and does not impact the conclusions that we will present later. \autoref{fig:merger-gas-star} shows the identified GMC and YMC candidates at a snapshot $t=0.25$~Gyr, after the first passage. Prominent tidal tails and bridges are developed at this epoch. Massive GMCs and young YMCs are not only formed in the central part of the two galaxies, but also broadly distributed along the tidal features.

We take advantage of the high cadence simulation output with a mean separation between adjacent snapshots around 1~Myr and track the orbits of all the YMCs that are formed at different epochs. We follow the \texttt{ParticleIDs} of all member stars within each YMCs across every simulation snapshots and construct a matching table to link YMCs in different snapshots. The tracking will be used to investigate the subsequent time evolution of the orbital and tidal information of YMCs in realistic galactic environments.

\section{Star and cluster formation during galaxy mergers}\label{sec:results}

\autoref{fig:merger-series} shows the time series of the simulated merger of two Milky Way-sized galaxies. The two galaxies first approach with each other for $\sim0.22$~Gyr until they experience the first passage. Immediately after the first interaction, prominent tidal tails appear at the tips of the two galaxies and extend to several tens of kpc. At $\sim0.52$~Gyr, the system experiences a second passage, which triggers a strong starburst that leads to large-scale gas outflows due to stellar feedback. At $\sim0.69$~Gyr, the final coalescence occurs. During the second passage and final coalescence, strong galactic torques are exerted onto the gas, reducing its angular momentum, and leading to significant gas inflow towards the center of the merger remnant.
In the following subsections, we will systematically investigate the effects of the major merger on star formation activities, focusing in particular on the properties of the ISM, GMC and YMC populations, the CFE, and the orbital and tidal properties of the YMCs.

\subsection{Enhancement of star formation during galaxy mergers}\label{sec:results-sfr}
\begin{figure}
\includegraphics[width=\columnwidth]{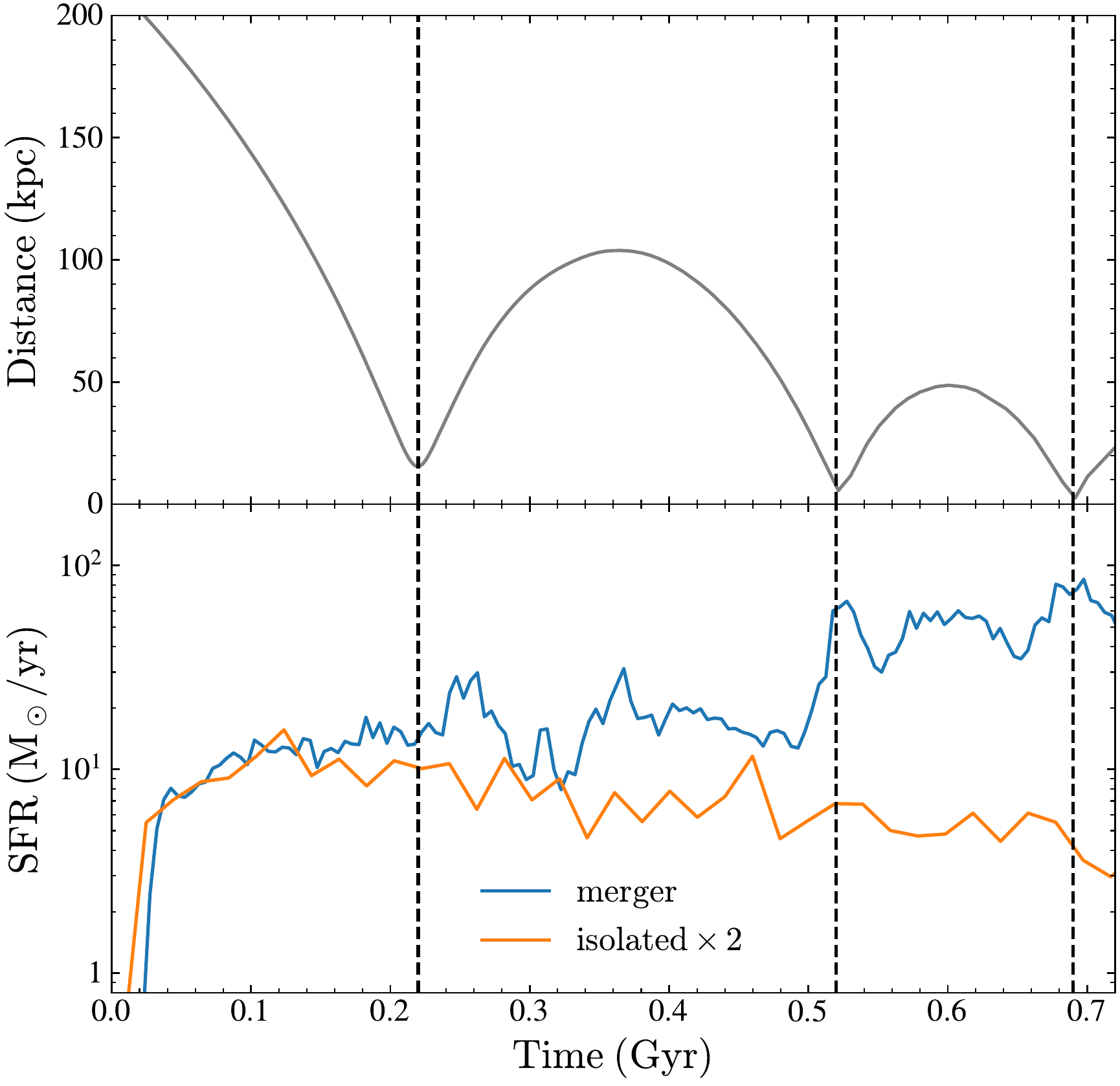}
\vspace{0mm}
\caption{Upper: Evolution of the distance between the centers of the two galaxies during the course of the simulation. Three close interactions at $\sim0.22$~Gyr, $\sim0.52$~Gyr, and $\sim0.69$~Gyr are labeled as vertical black dashed lines, repectively. Lower: Total SFR for the whole merger system (blue). For comparison, the SFR of the controlled simulation, an isolated Milky Way-sized galaxy, is also shown in orange. The SFR in the isolated case is multiplied by two for a fair comparison.}
  \label{fig:sfh}
\end{figure}

First, we examine the overall star formation history of the simulated merger throughout the whole merger sequence and compare it with the isolated galaxy case. The upper panel of \autoref{fig:sfh} shows the time evolution of the distance between the centers of the two galaxies for the merger orbit we described in \autoref{sec:methods-IC}. The center of each galaxy is determined by the center-of-mass of the bulge stars that are implemented in the initial conditions. The lower panel of \autoref{fig:sfh} compares the star formation history of the merger galaxy with that of an isolated case. We find that, as expected, galaxy mergers significantly boost the star formation rate (SFR) of the galaxies. A few tens of Myr after the first passage, the SFR of the merger reaches 20-30 $\Msun\,\text{yr}^{-1}$, two to three times higher than the isolated case. The rate stays at a high level even though the separation between the two galaxies increases.
When the two galaxies approach each other and interact again at $\sim0.52$ and $\sim0.69$ Gyr, the SFR skyrockets to a peak value approaching 100 $\Msun\,\text{yr}^{-1}$, which is more than ten times higher than in the isolated case.

It should be noted that the boosts of SFR for the first and last two interactions originate from different physical mechanisms. The star formation after the first passage is distributed broadly across the whole merging system, with intense star formation occurring not only in the nuclear regions of the two galaxies but also at the outskirts of the galaxies, the tidal tails and bridge. On the other hand, the second and final interactions are predominately nuclear starbursts produced by intense gas inflow towards the center of the merger remnant due to the significant angular momentum loss of the gas, \citep[see also][]{bournaud_etal11,renaud_etal19}.

\subsection{Enhancement of dense and high-pressure gas during galaxy mergers}\label{sec:results-gas}

\begin{figure}
\includegraphics[width=\columnwidth]{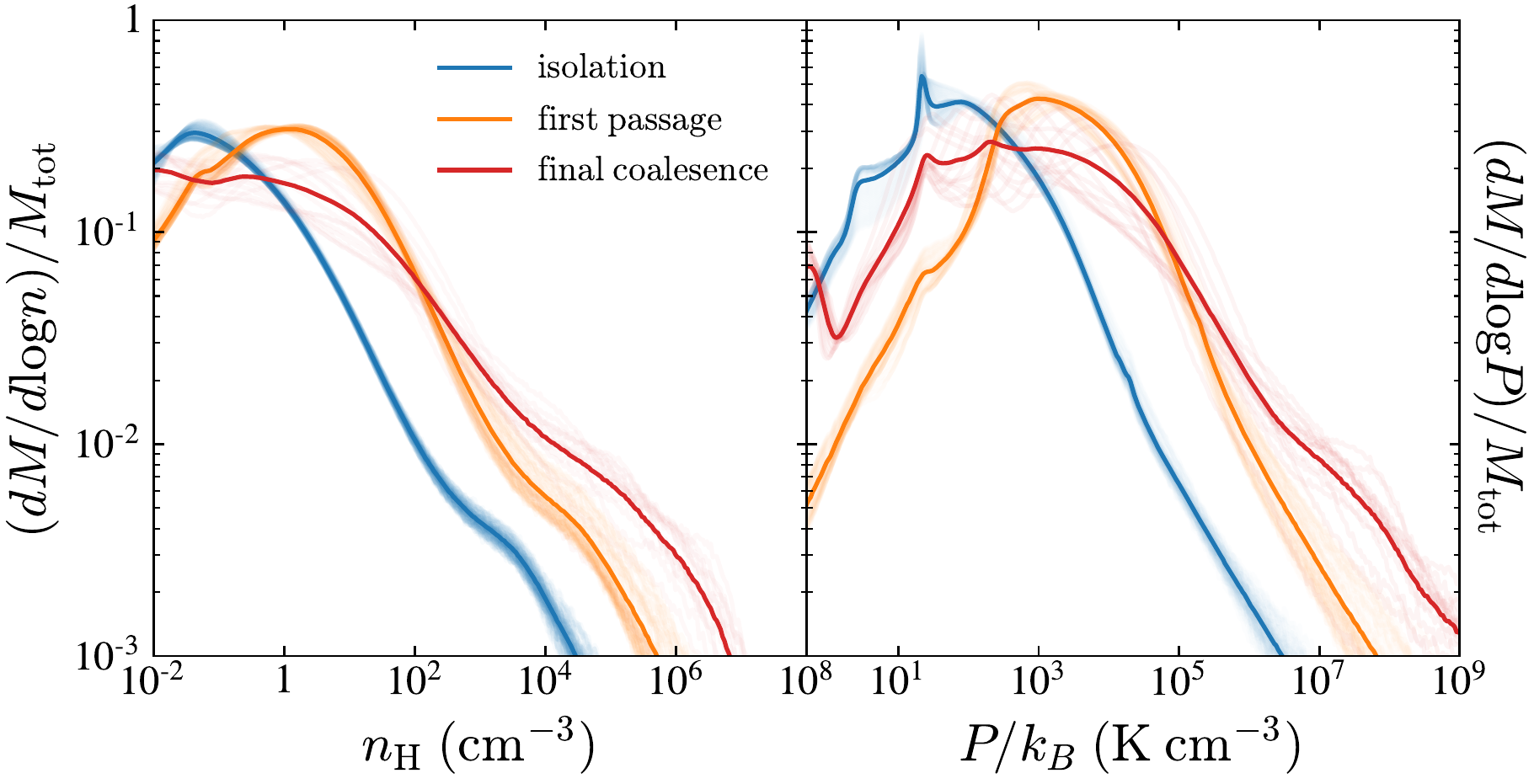}
\vspace{0mm}
\caption{Left: Mass-weighted PDF of gas number density in different situations: (1) in the controlled, isolated galaxy (blue), (2) during and right after the first passage (0.21-0.27~Gyr, orange); (3) during the final coalescence (0.65-0.72~Gyr, red). Right: the same as the left panel but for gas pressure. For both panels, thin lines represent the PDF derived from individual snapshots while the thick solid lines represent the median value of all available snapshots.}
  \label{fig:density-pdf}
\end{figure}

In addition to enhancing the SFR, we are particularly interested in how mergers alter the properties of the ISM and YMCs. In the following subsections, we compare various spatially-resolved properties in three different cases during the evolution of the simulated isolated and merging systems: (1) galaxies in the isolated simulations, (2) galaxies during the first passage (0.21-0.27~Gyr), and (3) galaxies during the final coalescence (0.65-0.72~Gyr). In this subsection, we focus specifically on the properties of the ISM in the simulated galaxies.

\autoref{fig:density-pdf} shows the mass-weighted probability density function (PDF) of the gas number density ($n_{\rm H}$) and gas pressure ($P/k_B$) in the three cases. The PDF of the gas number density (pressure) peaks at around 0.1 \rm cm$^{-3}$ (10 K cm$^{-3}$) and extends to $\sim10^4$ cm$^{-3}$ ($10^6$ K cm$^{-3}$) towards the high density/pressure ends when the galaxies are in isolation. Interestingly, both the PDFs of ($n_{\rm H}$) and $P/k_B$ show systematic shifts towards much higher values during the merger, suggesting that merger events trigger the creation of high-density and high-pressure gas, which is the ideal site for star and cluster formation.
For example, during the final coalescence, when a large fraction of the gas loses its angular momentum and flows into the central part of the merger remnant, the PDF of $n_{\rm H}$ ($P/k_B$) reaches $\sim 10^7{\rm cm}^{-3}$ ($10^9{\rm K \, cm}^{-3}$).

Note that the presence of the high density and pressure gas in the PDF depends sensitively on the choice of $\epsff$ in our simulations. As has been shown in L20, a higher $\epsff$ leads to a faster gas consumption timescale and the elimination of the highest density/pressure end of the PDF. However, the goal here is not to fully explore the sub-grid model parameters but to study the effects of mergers. Because, in this work, we design a control experiment that compares the simulated isolated galaxy and the merger with exactly the same galaxy formation model and parameters, the conclusion that mergers systematically enhance the formation of dense and high pressure gas is robust to the specific choice of model parameters, such as $\epsff$.

\subsection{Enhancement of the formation of the most massive GMCs and YMCs during galaxy mergers}\label{sec:results-massfunction}

\begin{figure}
\includegraphics[width=\columnwidth]{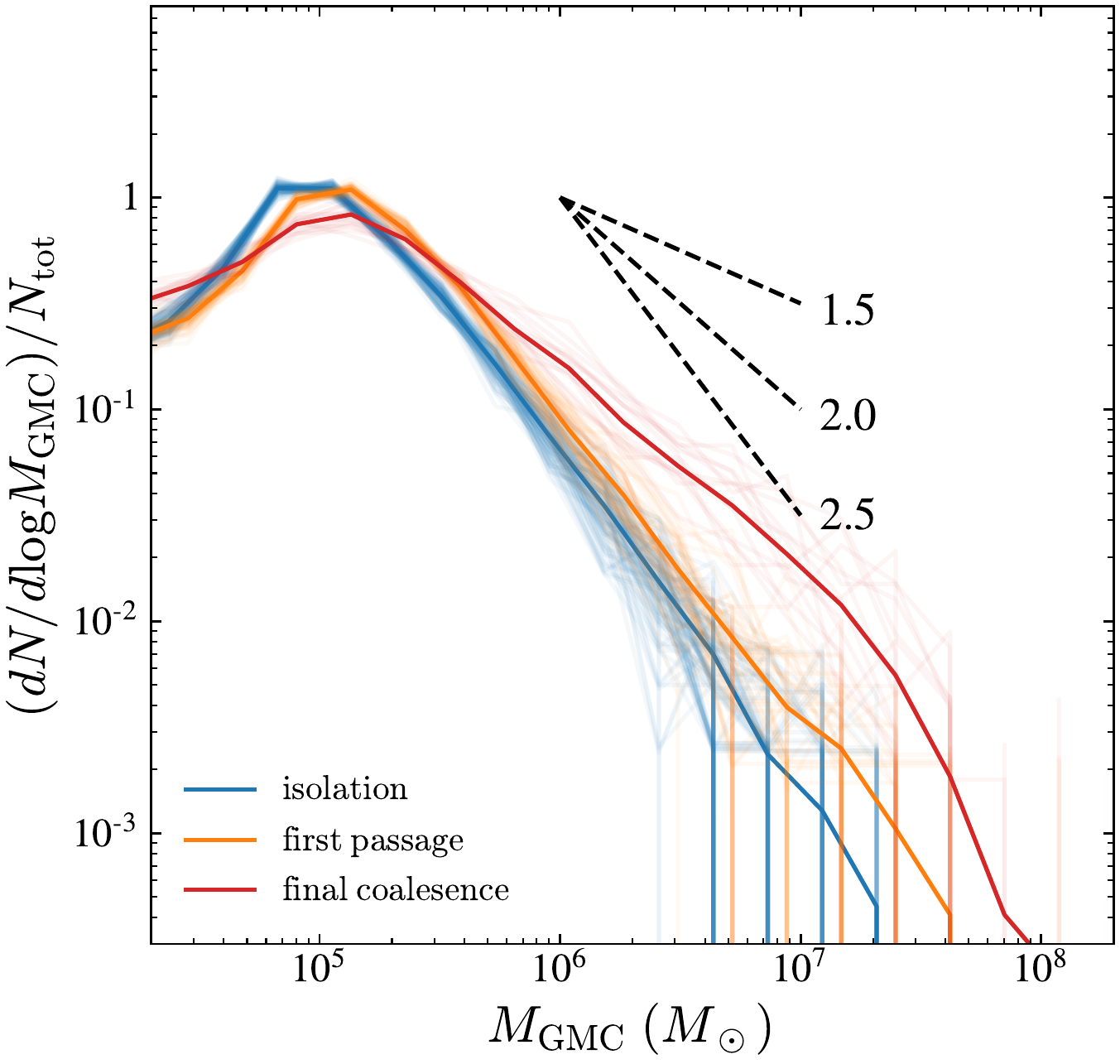}
\includegraphics[width=\columnwidth]{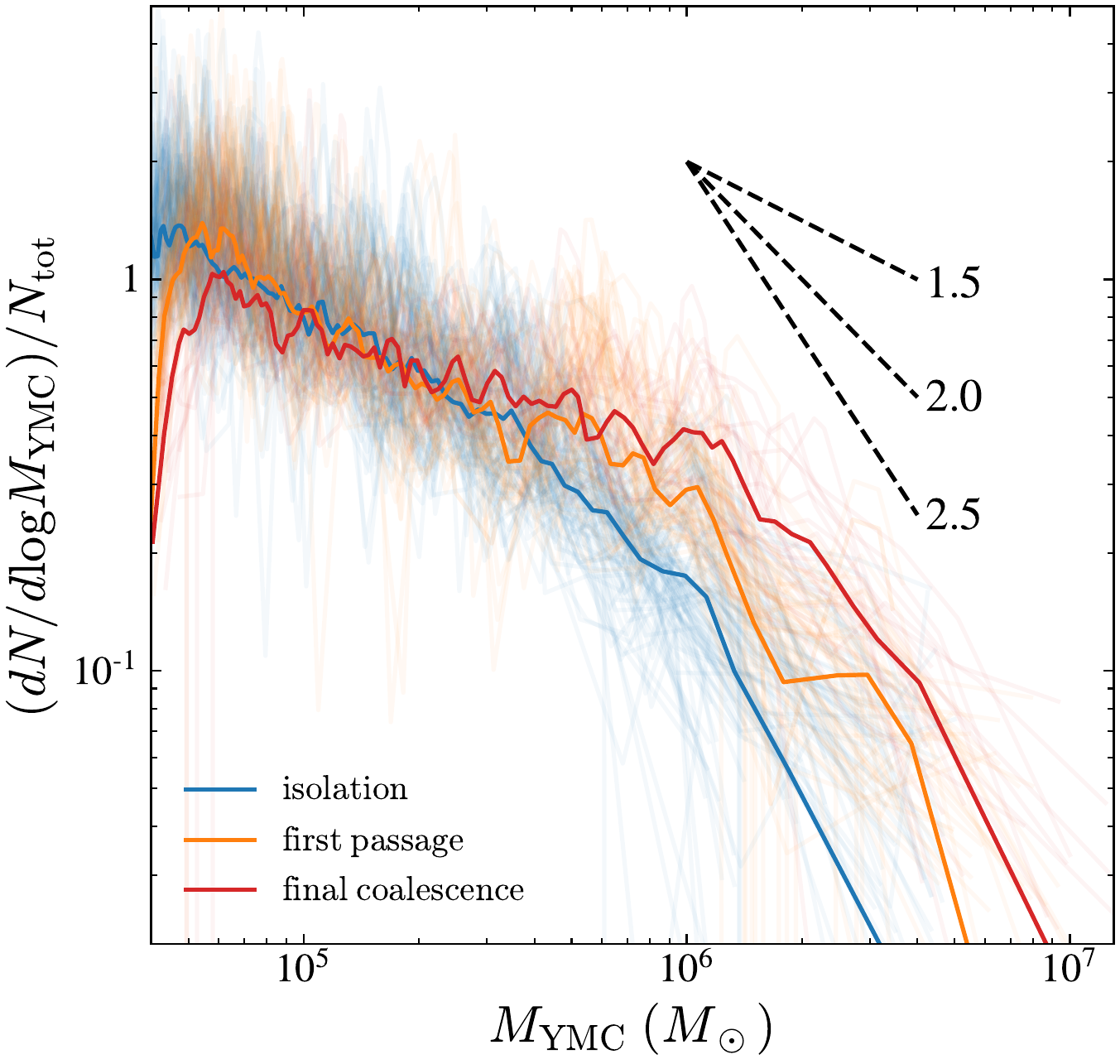}
\vspace{0mm}
\caption{Mass function of GMCs (upper) and YMCs (lower) in three different cases. GMCs/YMCs identified in controlled isolated galaxy, during first passage, and during final coalescence are labeled with a color style that is the same as \autoref{fig:density-pdf}. Only the YMCs that are younger than $<50$~Myr in each snapshot are used for the mass function calculations. For reference, three black dashed lines are overplotted to indicate power-law slopes of 1.5, 2.0, and 2.5 for each panel.}
  \label{fig:gmc-mass-function}
\end{figure}

We identify, at each simulation snapshot, bound GMC and YMC candidates using the methods described in \autoref{sec:methods-clustering}. GMCs are formed and disrupted continuously during the course of the simulations, therefore the GMC mass function reflects the instantaneous hierarchical structure of the cold phase ISM. In contrast, the identified YMC population contains clusters with all ages. Due to gravitational softening of the star particles and their large particle masses, we are unable to accurately follow the internal collisional dynamics and mass loss of the YMCs over a long period of time. Therefore, we focus only on YMCs younger than 50 Myr, which are not yet significantly affected by numerical artifacts.

\autoref{fig:gmc-mass-function} shows the mass function of GMCs and YMCs at three different evolution stages. Both the mass functions of the GMCs and YMCs can be described by a power-law distribution with a possible exponential cutoff at the high mass end.
To better quantify the shape of the mass function, we perform Bayesian power-law and Schechter function fits with a flat prior between -3 and -0.5 for the power-law slope and between $10^5$ and $10^9 \Msun$ for the cutoff mass.

The power-law slope of the GMC mass function is $2.47\pm0.02$, $2.32\pm0.01$, and $1.99\pm0.02$ for GMCs in isolated galaxies, during the first passage, and during the final coalescence, respectively\footnote{Keep in mind that the GMCs shown here are identified as gravitationally bound 3D objects. Therefore, the slope of mass function should not be used to directly compare with the observed 2D mass functions. We notice that the 3D GMC mass function is always steeper than those obtained in 2D (see L20).}. It is clear that galaxy mergers lead to a shallower GMC mass function and consequently the production of more GMCs with masses larger than $\sim10^6\Msun$. For example, the maximum GMC mass is $\sim10^7\Msun$ in isolated galaxies, while it can reach a few $10^8\Msun$ during the final coalescence. We also test whether there exists an exponential cutoff at the high mass end of the GMC mass function by performing a likelihood-ratio test. No statistically significant cutoff is found in all three cases with a $p$-value much smaller than 0.01.

Similarly to the GMC mass function, the YMC mass function also shows shallower power-law slopes during the merger. When a pure power-law fit is performed, the best-fitted slopes are $1.93\pm0.06$, $1.77\pm0.05$, and $1.67\pm0.02$ for YMCs formed in isolated galaxies, during first passage, and final coalescence, respectively. In contrast to the GMC mass functions, we find that YMC mass function for all three cases shows a statistically significant high-mass cutoff. Fitting a Schechter function, the slope (cutoff mass) becomes $1.55$, $1.50$, and $1.28$ ($2.51\times10^6\Msun$, $6.08\times10^6\Msun$, and $4.70\times10^6\Msun$) for the three cases, respectively. 
There exists a trend that a larger cutoff mass is obtained during mergers, especially in the final coalescence phase. Both the shallower mass function and higher cutoff mass during mergers lead to a significant enhancement of the formation of most massive YMCs during galaxy mergers. Encouragingly, similar phenomena have also been found in the most recent observations of nearby mergers from the HiPEEC survey \citep[][]{adamo_etal20}.
Generally speaking, the variation of the slopes and cutoff masses support a scenario in which the CIMF is not universal but changes dramatically with environment, consistent with recent observations of the shape of the CIMF in different types of galaxies \citep[e.g.,][]{johnson_etal17}.

\subsection{Enhancement of the cluster formation efficiency during galaxy mergers}\label{sec:results-cfe}

\begin{figure}
\includegraphics[width=\columnwidth]{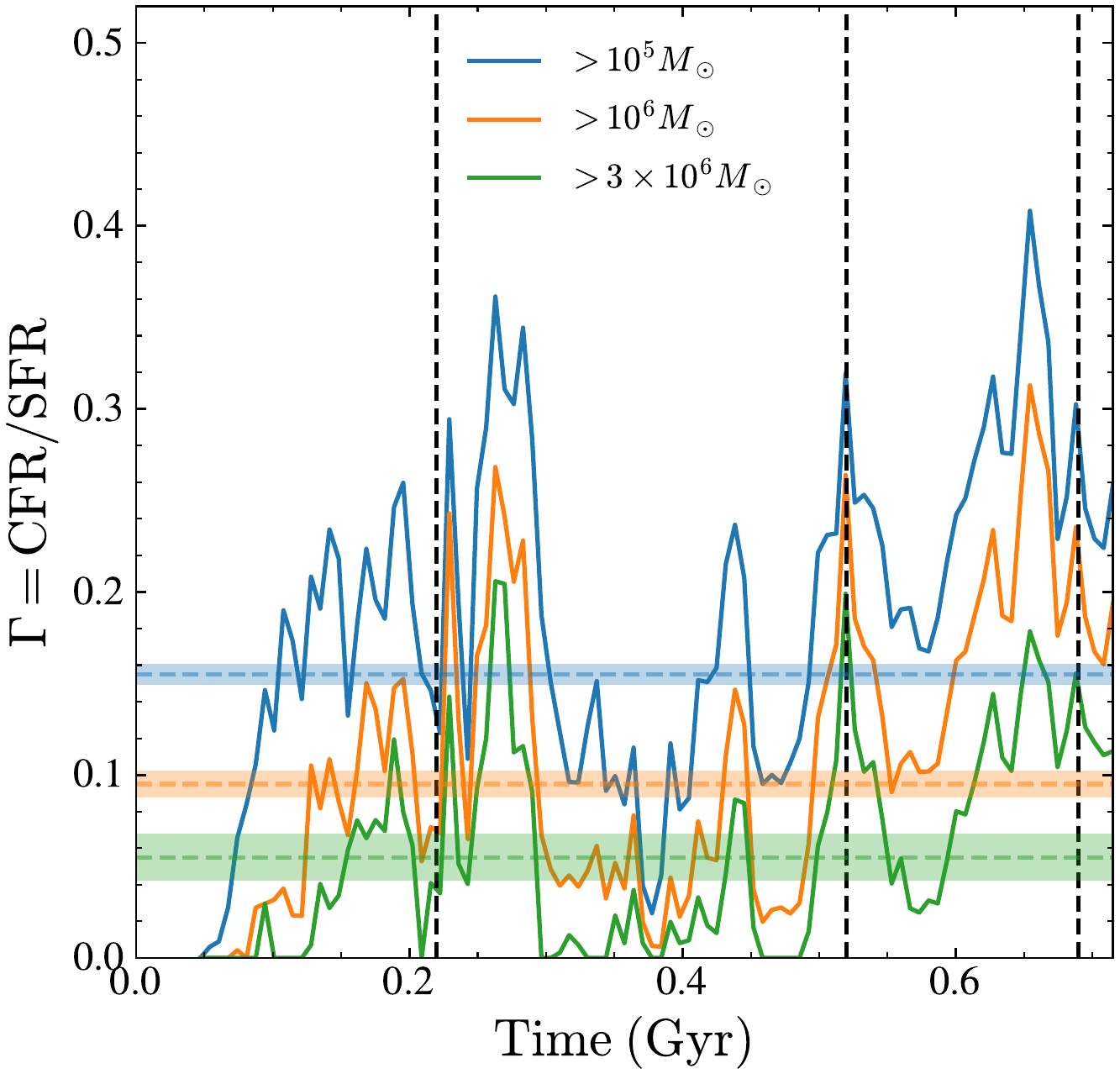}
\vspace{0mm}
\caption{Time-evolution of the CFE ($\Gamma$), defined as the ratio between the cluster formation rate and total SFR, during the course of merger. The efficiency is calculated for YMCs younger than 10~Myr and more massive than a threshold mass of $10^5\Msun$ (blue), $10^6\Msun$ (orange), and $3\times10^6\Msun$ (green). The median efficiencies in the isolated galaxy simulation using the same mass thresholds are shown as horizontal dashed lines. 
}
  \label{fig:CFE-evolve}
\end{figure}

\begin{figure*}
\includegraphics[width=2\columnwidth]{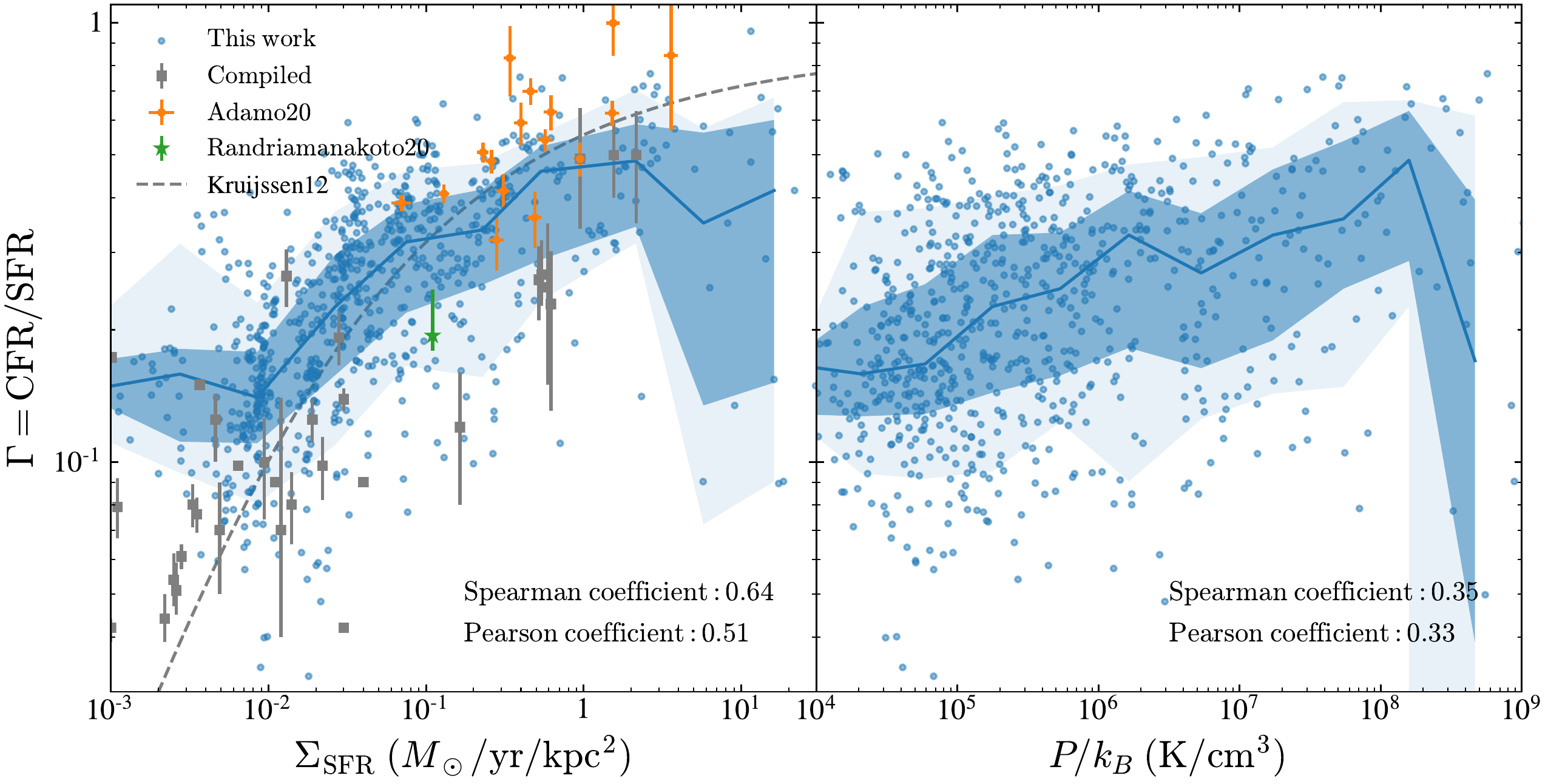}
\vspace{0mm}
\caption{Correlation between CFE ($\Gamma$) and SFR surface density ($\Sigma_{\rm SFR}$, left) and mass-weighted gas pressure ($P/k_B$, right). Blue dots are calculated in different galaxy patches from different simulation snapshots (see \autoref{sec:results-cfe-correlation} in detail). Light and dark blue shaded regions represent the 10-90 and 25-75 percentile ranges of the whole sample, respectively. Both the Spearman and Pearson linear correlation coefficients between $\log{\Ssfr}$ ($\log{P}$) and $\log{\Gamma}$ are shown in the lower right corner of each panel. Black and orange error bars in the left panel are observations in nearby disk galaxies and merging galaxies, respectively \citep[][]{adamo_etal15, randriamanakoto_etal19, adamo_etal20}. However, it should be noted that the absolute values of $\Gamma$ in observations depend on the extrapolation of mass function to lower mass, see \autoref{sec:results-cfe} for details. An analytical model of \citet{kruijssen12} between $\Ssfr$ and $\Gamma$ is overplotted as a dashed line.
}
  \label{fig:CFE-correlation}
\end{figure*}

One of the most important observables of YMC formation is the CFE, $\Gamma$, defined as the fraction of the total stellar mass forming in bound star clusters to the total stellar mass forming in the galaxy \citep[e.g.,][]{bastian_08}. In this subsection, we focus on the time evolution of the overall CFE for the entire merger system. In the next subsection, we will study the spatial variation of $\Gamma$ in different galactic environments.

To obtain $\Gamma$ in each simulation snapshot, we estimate the star and cluster formation rates by calculating the total mass of all stars and YMCs younger than 10~Myr and above a given mass threshold. \autoref{fig:CFE-evolve} shows the time evolution of $\Gamma$ in the merger simulation with three different YMC mass thresholds ($10^5$, $10^6$, and $3\times10^6\Msun$). For comparison, we calculate the median value of $\Gamma$ in the isolated galaxy simulation for the same thresholds to be 15.5\%, 9.5\%, and 5.5\%, respectively. First, we find that $\Gamma$ shows a large fluctuation over the course of the merger. The maximum $\Gamma$ can reach a value as high as 40\% for a threshold mass of $10^5\Msun$, suggesting that a large fraction of the star formation activities are contained in massive YMCs. In contrast, the median value of $\Gamma$ in the isolated disk simulations is only 15.5\%. The epochs of the peaks of $\Gamma$ clearly coincide with the pericentric passages and final coalescence, indicating that mergers strongly enhance CFE. This enhancement is 
the outcome of the changes of the shape of CIMF described in \autoref{sec:results-massfunction} and is a natural consequence of the formation of dense and high-pressure gas that triggers the formation of most massive GMC and YMC formation during galaxy mergers (see \autoref{sec:results-gas}). We emphasize that the enhancement mentioned here is for the cluster formation \textit{efficiency}, which is on top of the boost to the SFR during mergers in \autoref{fig:sfh}.

Note that the absolute value of $\Gamma$ we obtained here should not be used to directly compare with observations. Because of sensitivity limits, the minimum YMC mass that can be detected varies dramatically in different observations for different galaxies. In addition, different observations use different methods to take into account the missing low mass clusters by either applying a fixed low-mass cutoff or extrapolating the CIMF towards lower masses. Therefore, in this work, we simply highlight the \textit{relative} enhancement of $\Gamma$ during galaxy mergers in comparison to the values in isolated galaxies and  demonstrate that this enhancement is robust to the choice of different mass thresholds. 

\subsection{Spatial variation of cluster formation efficiency}\label{sec:results-cfe-correlation}

In the previous subsection, we studied the time evolution of the CFE, $\Gamma$, over the whole galaxy. Here we investigate how $\Gamma$ varies across different parts of the galaxies with different physical conditions. Because of the large variation of physical conditions during the course of the galaxy merger, our simulations are ideal to investigate the environmental-dependent cluster formation process.

In each snapshot, we split both galaxies into three concentric circular bins centered on the galaxy. The radius is determined so that each circle contains 20\%, 50\%, and 80\% of the stellar mass younger than 10~Myr. For each annulus, we calculate the SFR surface density ($\Ssfr$), the CFE ($\Gamma$), and the mass-weighted gas pressure $P/k_B$. We perform the analysis with different simulation snapshots and collect a large sample of galaxy patches with different environments.

\autoref{fig:CFE-correlation} shows the relationship between $\Gamma$ and $\Ssfr$ as well as with $P/k_B$.
Both $\Ssfr$ and $P/k_B$ vary over a wide range of values with $\Ssfr=10^{-3}-20\Msun\rm/yr/kpc^2$ and $P/k_B=10^4-10^9\rm K/cm^3$. The regions with the highest $\Ssfr$ or $P/k_B$ are typically the central regions of the galaxies during mergers.
Similarly, the efficiency $\Gamma$ changes dramatically as well, from $\sim1\%$ to near unity. Interestingly, there exists a strong positive correlation between $\Gamma$ and $\Ssfr$ (as well as $P/k_B$). This correlation is statistically significantly and is supported by both the Pearson (parametric) and Spearman (nonparametric) correlation coefficients between the logarithm of the quantities. These correlations support the environment-dependent cluster formation scenario and are consistent with previous observations in nearby late-type and merger galaxies \citep[e.g.][]{adamo_etal15,johnson_etal17,randriamanakoto_etal19,adamo_etal20}. They are also consistent with previous analytical works \citep[e.g.][]{kruijssen12} as well as numerical simulations \citep[e.g.][]{li_etal18, lahen_etal20}. 
The large scatter and a possible decline of $\Gamma$ at very high $\Ssfr$ ($P/k_B$) is mostly due to: (1) low number statistics because there are only a handful of annuli that reaches such extreme environment, and (2) the artificial disruption of YMCs during their very early evolution stage in dense and high-pressure regions.  

\section{Orbital and tidal history of star clusters}\label{sec:results-tracking}

\begin{figure}
\includegraphics[width=\columnwidth]{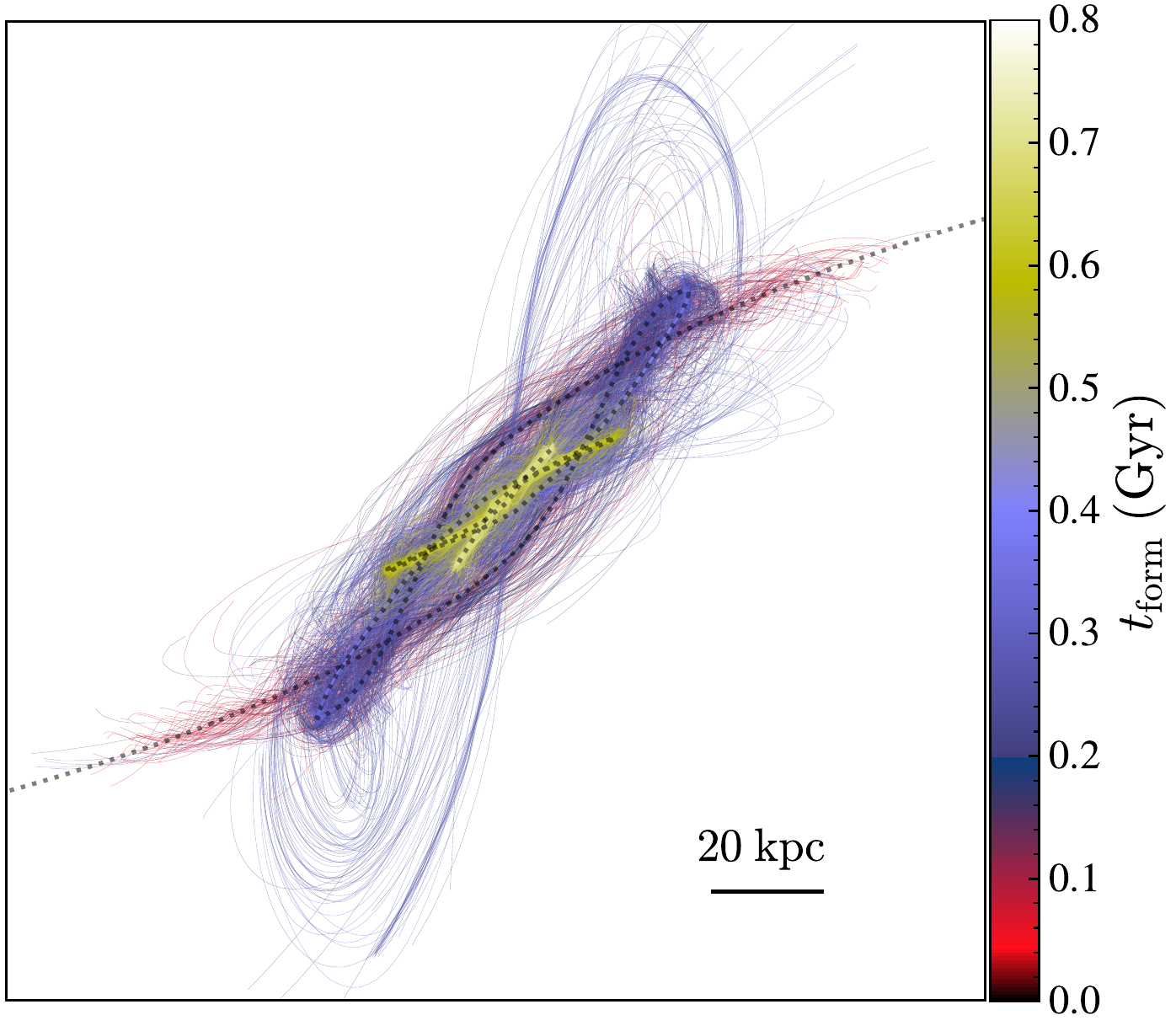}
\vspace{0mm}
\caption{Orbits of YMCs during the course of the merger color-coded by their formation epochs. For reference, black-dotted lines show the parabolic orbits of the centers of two galaxies. }
  \label{fig:cluster-orbit}
\end{figure}

In the last section, we focused on the formation of YMCs and its dependence on environment. We concluded that the formation efficiency of massive GMCs and YMCs is significantly enhanced during merger events. In this section, we focus on the subsequent evolution of YMCs after they emerge from their natal clouds.

We follow the orbits of each YMC from their formation to the end of the simulation. At each snapshot, we select all identified bound YMCs that are younger than 10~Myr and more massive than $10^5\Msun$. For each young YMC candidate, we record the \texttt{ParticleIDs} of all star particles that are members of the YMC and track the location of the YMC based on the recorded \texttt{ParticleIDs} in the subsequent snapshots. The center-of-mass of all of the star particles in the YMC is used to represent the position of the YMC. 
We notice that some YMCs are completely disrupted during the course of the simulations so that the star particles that originally belong to the YMCs are scattered over a large volume of space. We remove the disrupted clusters if their member stars have a spatial dispersion larger than 50~pc, a typical value of the tidal radius of globular clusters in our Galaxy. The number of removed clusters is around 10\% of total sample and does not affect our conclusions.

\autoref{fig:cluster-orbit} illustrates the orbits of all remaining YMCs that are formed at different epochs of the simulation. We find that YMCs that are formed before the first passage closely follow the orbit of their host galaxies. Similarly, YMCs that are formed during the second passage and final coalescence are located and remain at the very center of the merger remnant. On the other hand, a considerable fraction of YMCs formed during, or right after the first passage, have orbits that deviate strongly from their host galaxies, suggesting a rapid decoupling process between these YMCs and their host galaxies. We will quantify this decoupling and its consequence on their long-term dynamical evolution in \autoref{sec:results-evolution}.

Along their orbits, YMCs experience external tidal fields that determine their mass loss rate and survivability. Therefore, the tidal history of individual YMCs is crucial to understand the disruptive environments during mergers. To quantify the strength of the tidal fields, we calculate the tidal tensor along the trajectories of each YMC during the course of the simulations. The tidal tensor $\bm{T}_{ij}$ at position $\bm{x}_0$ under a time-varying gravitational potential field $\Phi(\bm{x},t)$ is defined as
\begin{equation}
    \bm{T}_{ij}(x_0,t)\equiv-\frac{\partial^2 \Phi(\bm{x},t)}{\partial x_i \partial x_j}\Biggr|_{\bm{x}=\bm{x}_0}.
\end{equation}
Following \citet{li_gnedin19}, $\bm{T}_{ij}$ is evaluated using a finite difference scheme centered at the position of the center-of-mass of the YMC over a scale of 50~pc. This scale is comparable to estimates of the tidal radii of the Galactic GCs and is also consistent with previous practice \citep[e.g.][]{renaud_etal17}. We use the maximum of the absolute value of the eigenvalues $\ltid$ of the tensor $\bm{T}_{ij}$  as a proxy of the tidal strength.
Moreover, we include the term $-1/3\sum_i\lambda_i$ that represents the Coriolis force \citep[e.g.][]{renaud_etal11, pfeffer_etal18}. We find that this term is in general not important for most YMCs, but can significantly reduce the tidal effect when the YMCs sink into the very center of the merger remnant, a result that is consistent with \citet[][]{pfeffer_etal18,keller_etal20}. Quantitatively, the median $\lambda$ decreases from $8\times10^7$ to $3\times10^7 {\rm Gyr}^{-2}$ for YMCs within the central 1~kpc of the merger remnant.
In the next two subsections, we will separately examine the short- and long-term evolution of the tides experienced by YMCs formed in different galactic environments and at different merger stages.

\subsection{Tidal disruption in the early stage depends on cluster mass}\label{sec:results-tides-mass}

\begin{figure}
\includegraphics[width=\columnwidth]{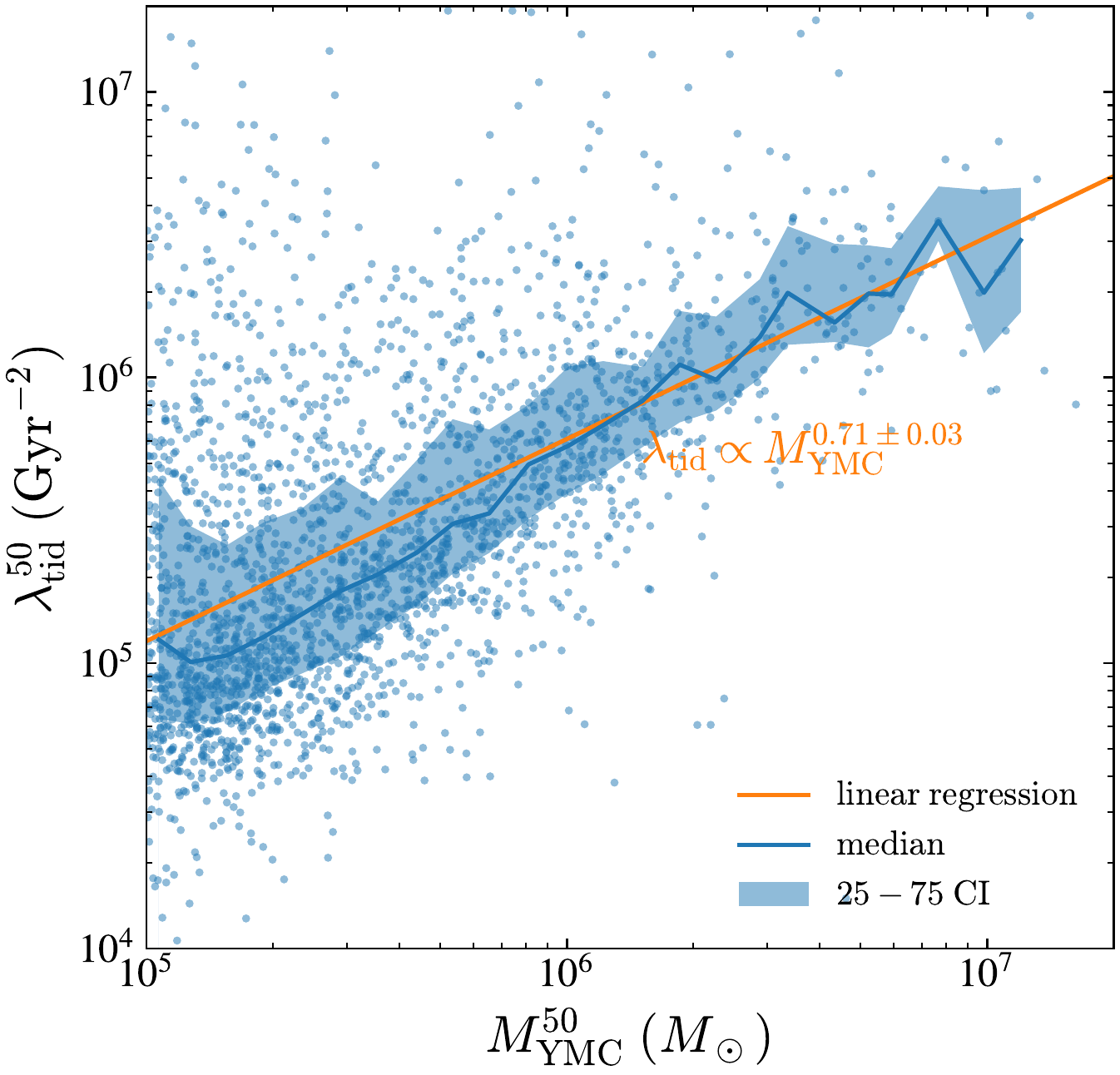}
\vspace{0mm}
\caption{Initial YMC mass vs. the median tidal strength $\ltid$ experienced by the YMCs during the first 50~Myr after their formation. Blue solid lines and shaded regions represent the median relation and the 25-75 percentile range, respectively. The best-fitted linear relation between $\log{M_{\rm YMC}}$ and $\log{\ltid}$ is also overplotted as an orange line.
}
  \label{fig:cluster_tidal-mass}
\end{figure}

In \autoref{sec:results-massfunction}, we showed that mergers produce a large number of YMCs over a wide mass range from a few times $10^4$ to $10^7\Msun$. This allows us to examine whether there exists any mass-dependence of the strength of the tidal field around each YMCs in their early evolution stage. For all YMCs more massive than $10^5\Msun$, we track the time evolution of $\ltid$ for the first 50~Myr after their formation and calculate its median value during this period, $\ltidfifty$. \autoref{fig:cluster_tidal-mass} shows the relationship between $\ltidfifty$ and the mass of the YMCs, $M_{\rm YMC,50}$.
Because YMCs first gain mass via accretion during the early formation phase before emerging from their natal clouds and then lose mass due to dynamical disruption, we define $M_{\rm YMC,50}$ as the maximum mass that the YMC has ever reached within the first 50~Myr of their lives.

First, we find that the tidal strength for most YMCs is quite large ($\ltidfifty\gg10^4{\rm \, Gyr}^{-2}$), in comparison to the typical value of the tidal fields produced by the large-scale gravitational field of the Galaxy ( $\bar{\lambda}_\odot\approx1600{\rm \, Gyr}^{-2}$, see \citealt[][]{renaud_etal17}). This means that the tidal field experienced by the YMCs right after their emergence from their natal clouds depends on their interaction with local concentrations of dense gas. It is the small-scale structure of the ISM, rather than the large-scale smooth galactic tides, that dominates the dynamical disruption process \citep[][]{gieles_etal06, kruijssen_etal11} in the early stage.

More interestingly, we find a strong positive correlation between $M_{\rm YMC,50}$ and $\ltidfifty$ with the best-fit relation between the two quantities given by:
\begin{equation}
    \ltidfifty\propto M_{\rm YMC,50}^{0.71\pm0.03}.
\end{equation}
This positive correlation suggests that more massive clusters tend to reside in regions with higher tidal fields immediately after they form. It is understandable because of the hierarchical nature of the ISM: YMCs with higher masses are preferentially formed from higher overdensity regions that assemble much more cold and dense gas than YMCs with lower masses. The clumpiness of the overdensity around the massive YMCs provides strong tidal fields that lead to rapid disruption in the early stage.

Previous N-body simulations of clusters in tidal fields showed that the disruption timescale can be expressed as a function of cluster mass and tidal field strength: $\ttid(M) \propto M^{2/3}\ltid^{-1/2}$
\citep[e.g.][]{baumgardt_makino03,gieles_baumgardt08}. When $\ltid$ is not a function of cluster mass, the disruption timescale is $\ttid(M)\propto M^{2/3}$, a scaling that is a commonly used when calculating the time evolution of cluster populations. However, as described above, $\ltid\propto M^{0.71}$ is a strong function of cluster mass in the early evolutionary stage and therefore alter the mass-dependency of tidal disruption timescale to $\ttid\propto M^{2/3}\ltid^{-1/2}\propto M^{0.31}$. This change has obvious effects on the evolution of the mass function and age distribution of YMCs, which can potentially be quantified by future observations of YMC populations in nearby galaxies.
In Appendix~\ref{sec:discussion-tides}, we derive a set of analytical expressions that describe quantitatively how different mass-dependent tidal strengths affect the time evolution of the mass function of clusters in both the impulsive and continuous YMC formation cases.

\begin{figure*}
\includegraphics[width=1.8\columnwidth]{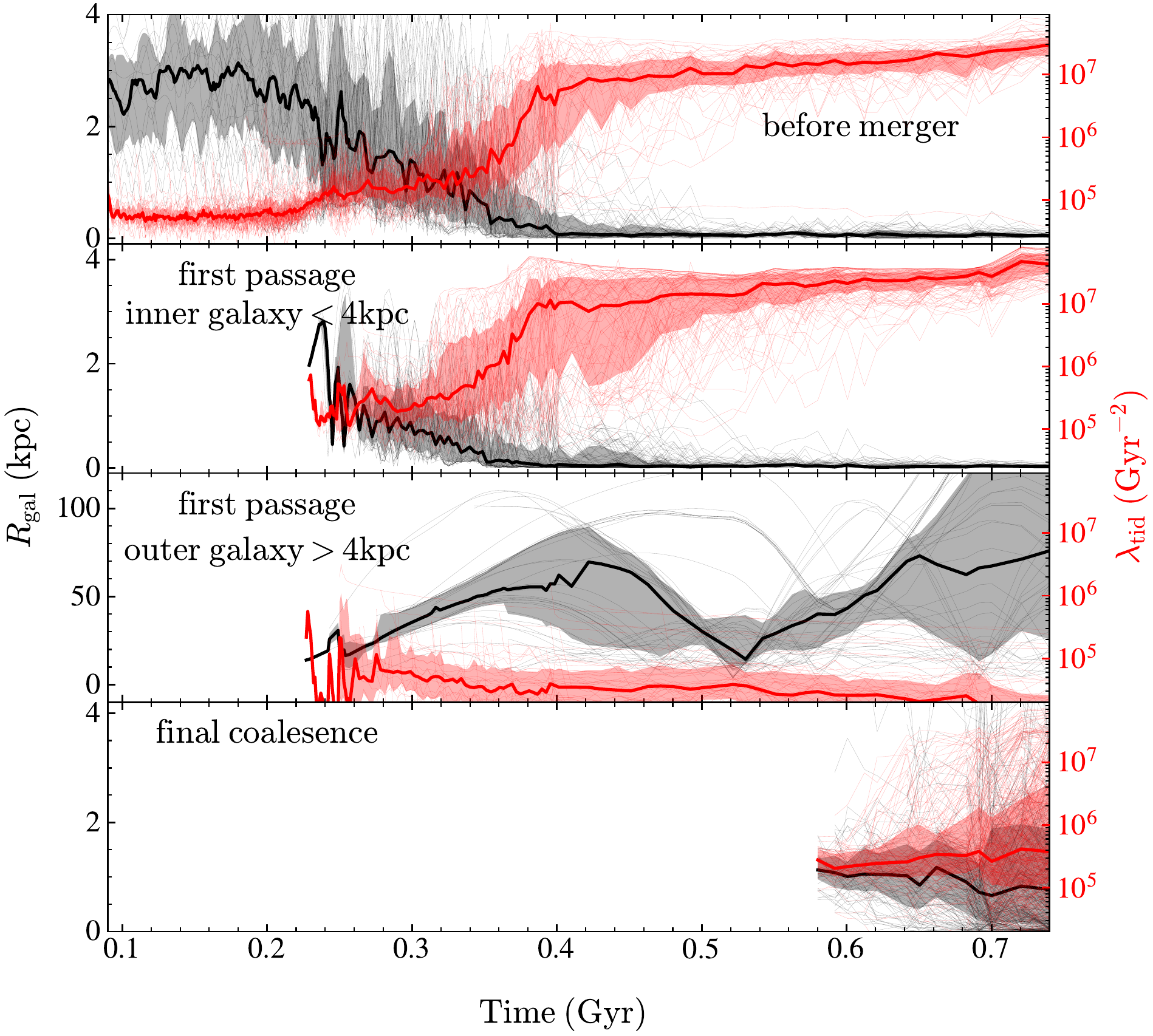}
\vspace{0mm}
\caption{Tracking the orbital and tidal information of YMCs during the course of the merger. The time evolution of the galacto-centric radius ($\rgal$) and the maximum eigenvalue of the tidal tensor ($\ltid$) of individual YMCs are plotted as thin black and red lines, respectively. Thick solid lines and shaded regions represent the median and 25-75 percentile range of the entire YMC sample. Different panels show the statistics of YMCs formed at different epochs and different locations: from top to bottom, 1) before the first passage ($t_{\rm form}<0.2$~Gyr), 2) during the first passage ($t_{\rm form}=[0.22-0.3]$~Gyr) and within the inner region of the galaxies ($\rgal<4$~kpc), 3) during the first passage ($t_{\rm form}=[0.22-0.3]$~Gyr) and in the outer galactic regions ($\rgal>4$~kpc), and 4) during the final coalescence ($t_{\rm form}>0.58$~Gyr).}
  \label{fig:cluster-orbit-tides}
\end{figure*}

\subsection{Long-term evolution YMCs across different merger stages}\label{sec:results-evolution}
To better investigate how mergers affect the orbital and tidal properties of YMCs, we plot the time evolution of the galacto-centric radii of all YMCs in \autoref{fig:cluster-orbit-tides}. We split the whole sample into four groups based on their formation epoch and location: YMCs formed (1) before first passage, (2) during the first passage and within the inner 4~kpc of their host galaxies, (3) during the first passage and in the outer galaxies with $\rgal>4$~kpc (including tidal tails and the bridge), and (4) during the final coalescence. In agreement with the qualitative impression shown in \autoref{fig:cluster-orbit}, we find that YMCs formed before the first passage quickly sink into the center of the galaxies in about $0.4$~Gyr. 
The spiral-in process is caused by dynamical friction against the ISM and field stars. The effect of dynamical friction is significantly enhanced due to the extremely large orbital eccentricities caused by first passage at $0.22$~Gyr. We investigate the orbits of these clusters and find that their initial orbits are mostly circular with orbital eccentricities $e<0.2$. After the first passage, the orbits of a large fraction of the clusters become highly eccentric with $e>0.8$. The large eccentricity reduces the dynamical friction timescale significantly \citep[e.g.][]{lacey_cole93,vandenbosch_etal99}, which leads to rapidly shrinking orbits. Similarly, YMCs formed in the inner galaxy during the first passage follow similar trends as they quickly sink into the galaxy centers before $0.5$~Gyr. On the other hand, a large fraction of the YMCs that are formed in the outer regions of the galaxies during the first passage are ejected from their host galaxies with $\rgal$ reaching 50-100~kpc. Many of them are formed in the tidal tails and bridge with large initial angular momentum. Their interaction with the merger remnant leads to further ejection. Similar behavior of cluster ejection in mergers has been discussed in previous theoretical works and numerical simulations \citep[e.g.][]{kruijssen_etal12, renaud_gieles13} and is believed to be the one of the key mechanisms leading to the long-term survival of globular clusters \citep[e.g.][]{kravtsov_gnedin05, kruijssen15}.

In addition to $\rgal$, \autoref{fig:cluster-orbit-tides} also shows the time evolution of $\ltid$ for the same four groups of YMCs. Immediately after their formation, YMCs in different groups experience similar tidal fields with $\ltid\sim10^5-10^6 {\rm Gyr}^{-2}$, suggesting that the tide fields around YMCs in the early stage depend predominately on the properties of the surrounding medium and are not sensitive to different evolutionary stages of the galaxy merger. The subsequent evolution of $\ltid$, however, is largely controlled by the YMC orbits. For YMCs in the first and second groups, the tidal strength increases significantly to extremely large value, $\ltid>10^7{\rm Gyr}^{-2}$, when they sink into the center of the merger remnant after $t=0.4$~Gyr. They are quickly disrupted due to the strong tides and eventually join the bulge or even the nuclear star cluster of the merger remnant. Moreover, YMCs formed during the final coalescence are mostly located in the innermost region of the galaxies. These YMCs will suffer the same fate as the above two groups, sinking into the center of the remnant and disrupting. 
In contrast, YMCs in the third group show a decreasing $\ltid$ because of the increasing $\rgal$ after their formation.  Roughly 0.3~Gyr after their formation, these YMCs are ejected from the dense gaseous disk into the galaxy halo. When this happens, $\ltid$ decreases to less than $10^4 {\rm Gyr}^{-2}$. When the simulation stops at $t\sim0.75$~Gyr, most of these YMCs are still gravitationally bound, demonstrating the possibility that they might survive for long cosmic times and may be promising galactic GC candidates.

\section{Discussion}\label{sec:discussion}
\subsection{Comparison to previous works}\label{sec:discussion-compare}

Over the last decade, there have been a few numerical simulations that focused on investigating the effects of galaxy mergers on star cluster formation. Here we compare our findings with some of those studies with similar setup.

As we have shown in \autoref{sec:results}, mergers not only increase the global SFR, but also trigger the formation of the most massive star clusters and enhance the CFE. 
This enhancement is clearly associated with the enhancement of cold dense molecular gas with the highest pressures, a condition that has been explored by previous simulations \citep[e.g.][]{renaud_etal15, li_etal17, li_etal18, ma_etal20, lahen_etal20} and analytical models \citep[e.g.][]{kruijssen12}.
For example, \citet[][]{renaud_etal15} found a significant enhancement of both star and cluster formation rate in a parsec-resolution simulation of Antennae galaxies, in agreement with our results. The CFE during the first passage of the merger also increases significantly. One distinct difference between their results and ours appears when the two galaxies reach the final coalescence phase. While the overall SFR is significantly enhanced in their simulation, only a small fraction of the SFR is contributed to clusters. The star formation activities during the final coalescence are limited to a compact region at the very center of the merger remnants, suggesting that gas is transferred to the galactic center very quickly before fragmenting into smaller pieces. In contrast, our current work presents an alternative picture in which the merger remnant during the final coalescence is very turbulent and clumpy, therefore generating a huge number of YMCs, whose masses dominate the total SFR. This difference possibly arises from different cooling/heating prescriptions that control gas fragmentation, different feedback mechanisms that are responsible for turbulence injection, or different gravity solvers for star particles \footnote{The force resolution for star particles is adaptive and depends on the local grid size in \textsc{RAMSES}, but is a fixed value in \textsc{arepo}.} that control the dynamics of the YMCs. 

Most recently, \citet[][]{lahen_etal20} simulated the merger of two dwarf galaxies with much higher resolution, so that the formation and evolution of individual massive stars are resolved. Similar to our work, this simulation produced many star clusters during the final coalescence. Future works with varying numerical implementations are needed to investigate the origin of such discrepancies.
In either case, we note (as described in \autoref{sec:results-evolution}), the YMCs formed during the final coalescence sink into the center of the merger remnant and get disrupted very quickly due to extremely strong dynamical friction and tidal forces. Therefore, in the long term, it appears not to be important whether star formation in the central coalescence happens preferentially in clusters or not.

Regarding the CIMF, in \autoref{sec:results-massfunction}, we show that the CIMF in different stages can be described by a power-law. A power-law CIMF is commonly seen in previous simulations \citep[e.g.][]{kravtsov_gnedin05,bournaud_etal08,powell_etal13}. We also find that the slope of the CIMF is shallower during merger events than in the isolated case, which has also been seen in previous isolated \citep[e.g.][]{lahen_etal20} and cosmological simulations \citep[e.g.][]{li_etal17,li_etal18, ma_etal20}. This result seems to be intriguing because it is not consistent with the well-known power-law slope of -2 for the observed CIMF in the Antennae galaxies \citep[e.g.][]{zhang_fall99}. However, we argue that the Antennae are a specific pair of galaxies in a particular merger phase. It is not clear whether the universality of the slope of the CIMF can be generalized to all galactic environments. Encouragingly, most recently spatially-resolved observations, such as the HiPEEC \citep[][]{adamo_etal20} and GOALS \citep[][]{linden_etal21}, found clear evidence of shallower slopes of the CIMF during mergers, supporting the conclusions in this work. More research, both in observations and simulations, is needed to investigate the shape of the CIMF in different galactic environments.

\subsection{Connection to the origin of globular clusters}
 
Since the HST discoveries of abundant YMCs in interacting galaxies, such as NGC 1275 \citep[e.g.][]{holtzman_etal92} and Antennae \citep[e.g.][]{whitmore_etal99}, galaxy mergers have been considered as the most efficient factory of YMCs. Naturally, globular clusters, whose mass and size are similar to YMCs, are hypothesized to be formed during these events as well \citep[e.g.][]{ashman_zepf92}.
In this paper, using hydrodynamical simulations that resolve the formation of individual YMCs, we show that galaxy mergers systematically alter the shape of the YMC mass function, resulting in a shallower mass function and higher cutoff mass during mergers. These changes lead to a significant enhancement of the formation efficiency of YMCs during mergers.
However, we emphasize that this is not to say that galaxy mergers are the dominant channel for the formation of globular clusters. Answering the latter question is a much harder task and clearly beyond the scope of this paper.

Recent observations of globular clusters, such as the tight linear correlation between the total mass of the GC system and host galaxy halo mass \citep[e.g.][]{spitler_forbes09, harris_etal17b} and the broad range of GC metallicity distribution \citep[e.g.][]{peng_etal06, harris_etal06, usher_etal12}, suggest that GCs trace well the mass assembly and metal enrichment history of their host galaxies \citep[][]{kruijssen_etal20}. Therefore, the co-evolution of GCs and galaxies can only be understood within the framework of hierarchical structure formation, making the origin of globular clusters a multi-scale, multi-physics problem.
Specifically, whether mergers are the main formation channel is still subject to hot debate. One the one hand, as is shown in this paper and other previous works, gas-rich major mergers significantly enhance the formation efficiency of GCs. This merger-induced GC formation scenario has been adopted to build several semi-analytical models that successfully reproduce many GC-related scaling relations \citep[][]{muratov_gnedin10,li_gnedin14, choksi_etal18, choksi_gnedin19}. One the other hand, major mergers are rare events with an active phase that only lasts for a small fraction of cosmic time \citep[e.g.][]{keller_etal20}. They are also thought to preferentially disrupt YMCs that sink into the center of the merger remnants \citep[e.g.][]{kruijssen_etal11, rieder_etal13}. Though huge efforts have been made in recent years \citep[e.g.][]{li_etal17,li_etal18,li_etal19, pfeffer_etal18, reina-campos_etal22}, to quantitatively answer this question requires comprehensive modeling of the formation, orbital and dynamical evolution of YMCs in hierarchical structure formation.

\subsection{Limitations of this work}

We note that there are a few limitations of this work. First, the experiment we use here relies on a merger simulation of two Milky Way-sized galaxies under a progradge-prograde configuration, which provides the maximum effects of tidal interaction and highest enhancement on star cluster formation. Different configurations of mergers and different properties of host galaxies are needed to fully explore the parameter space \citep[e.g.][]{bournaud_etal05, cox_etal08, lotz_etal10gas,lotz_etal10mass,tress_etal20}.
Second, although our simulations have high mass ($\sim10^3\Msun$) and spatial ($\sim3$~pc) resolutions in the standard of galaxy simulations, they are far from resolving the internal structure of the individual star clusters. Therefore, the long-term dynamical evolution, i.e. the mass loss due to tidal disruption, of the YMCs identified in the simulations is not properly modeled. We therefore cannot investigate the detailed dynamical evolution of star clusters after they emerge from their natal clouds. We can only treat cluster particles as tracers to record the time-evolution of their tidal environment \citep[see also][]{renaud_etal17}.
Finally, several physical ingredients, such as magnetic fields and run-time radiative transfer, is not implemented in \texttt{SMUGGLE}. These missing physics may potentially affect the properties GMC and YMC populations and will be explored in future works (Li et al. in preparation).

\section{Summary}\label{sec:summary}
We performed a high-resolution hydrodynamic simulations of a merger of two Milky Way-sized galaxies under the framework of the \textsc{Arepo}-\texttt{SMUGGLE} model. We design a controlled experiment by comparing the merger simulation with an isolated galaxy (L20) and investigate how galaxy mergers affect the structure of the ISM and the properties of the GMC and YMC populations. We find that galaxy mergers profoundly change the cold and dense phase of the ISM and trigger an efficient formation of the most massive GMCs and YMCs. Below we list our main findings.

\begin{itemize}
    \item The SFR of the simulated galaxy merger reaches $30\Msun\, \text{yr}^{-1}$ during the first passage, about three times higher than the rate for the two galaxies in isolation. After the second passage, a large fraction of gas loses its angular momentum, sinks into the nuclear region of the galaxy, and triggers a significant nuclear starburst, keeping an extreme SFR above $50\Msun\, \text{yr}^{-1}$ until the simulation stops after the final coalescence.
    \item The mass-weighted PDF of both gas density and pressure extends to much higher values in all merger events compared to the isolated case, suggesting that merger-induced galactic tides effectively compress gas to high density and high pressure. Gas pressure in the tidal tails and centers of galaxies during mergers reaches $10^7-10^9\rm K/cm^3$, several orders of magnitude higher than the value in the isolated galaxy.
    \item The mass function of GMCs above the resolution limit is best described by a pure power-law without high-mass cutoff. Comparing to GMCs in isolated galaxies, the mass function has a shallower power-law slope during mergers and extends to higher GMC masses. Similarly, the mass function of YMCs also shows a power-law shape with a shallower slope during mergers ($\sim-1.7$) compared to the slope in the isolated case ($\sim-2.0$).
    \item In contrast to the GMC mass function, the CIMF shows a statistically significant high-mass exponential cutoff. By fitting a Schechter function, we find that the cutoff mass of the YMC mass function is systematically higher during mergers ($5-6\times10^6\Msun$) than in the isolated case ($\sim2.5\times10^6\Msun$). Both the shallower slopes and higher cutoff mass of the CIMF are consistent with the most recent HiPEEC survey \citep[][]{adamo_etal20} and luminous infrared galaxies in GOALS \citep[][]{linden_etal21}.
    \item We calculate the time evolution of the CFE and find that this efficiency is strongly enhanced during all merger events. We emphasise that the enhancement mentioned here is for the efficiency, which is on top of the boost in the overall SFR during mergers.
    \item We spatially partition the galaxy pair into several galacto-centric annuli, calculate the spatially-resolved CFE and compare it with the SFR surface density $\Ssfr$ and mass-weighted gas pressure $P/k_B$. Consistent with observations and previous theoretical models, we find a strong positive correlations between $\Gamma$ and $\Ssfr$ as well as $P/k_B$. This correlation demonstrates the environment-dependent cluster formation scenario in a spatially-resolved fashion.
    \item We follow the trajectories of the model YMCs formed at different epochs to investigate their orbital motions and tidal field experienced during the mergers of their host galaxies. We find that majority of the YMCs formed before the first passage and during final coalescence sink into the center of the merger remnant very efficiently due to dynamical friction.
    These clusters experience strong tidal fields with the maximum eigenvalue of the tidal tensor $\ttid>10^8\,{\rm Gyr}^{-2}$ and are quickly disrupted as they spiral in. On the other hand, a large fraction of YMCs formed during the first passage are re-distributed to much larger galacto-centric radii and therefore experience much weaker tidal disruption.
    \item With a broad mass range of model YMCs formed in the simulations, we find a strong mass-dependent tidal field around the YMCs in the first $\sim50$~Myr of their life: $\ltid\propto M_{\rm YMC}^{0.71\pm0.03}$. This mass-dependency propagates to the tidal disruption timescale and alters the dependence between the timescale and cluster mass ($\ttid\propto M^{2/3}\ltid^{-1/2}\propto M^{0.31}$). This new correlation leads to a new scaling on the evolution of the mass and age distribution of the YMC population in the early stage.
    \item Our simulations demonstrate that galaxy mergers significantly enhance the formation efficiency of most massive star clusters, the most promising candidate of GCs. YMCs that are formed pre-merger or in the latest final merger stages quickly sink into the center of the merger remnants and are efficiently disrupted. In contrast, YMCs formed at the first passage of the galaxy mergers when a large-scale tidal structures are easily ejected to the outer halo and potentially survive the subsequent dynamical evolution over cosmic time. These YMCs are likely to survive to the present and become promising candidates for globular clusters.
\end{itemize}

\section*{Acknowledgements}
It is our pleasure to thank the referee, Diederik Kruijssen, for his very constructive comments and suggestions, which greatly improved the manuscript.
We thank Volker Springel for giving us access to \textsc{AREPO}.
We are grateful to Oleg Gnedin, Mordecai-Mark Mac Low, Jeremiah Ostriker, Aaron Smith, and Vadim Semenov for insightful comments and suggestions.
HL was supported by NASA through the NASA Hubble Fellowship grant HST-HF2-51438.001-A awarded by the Space Telescope Science Institute, which is operated by the Association of Universities for Research in Astronomy, Incorporated, under NASA contract NAS5-26555. 
FM acknowledges support through the program "Rita Levi Montalcini" of the Italian MUR. 
MV acknowledges support through NASA ATP grants 16-ATP16-0167, 19-ATP19-0019, 19-ATP19-0020, 19-ATP19-0167, and NSF grants AST-1814053, AST-1814259, AST-1909831, AST-2007355 and AST-2107724.
GB acknowledges support from HST-AR-15022, NSF grant OAC-1835509, a NASA TCAN grant, as well as the Simons Foundation.
LVS acknowledge support from the NASA ATP80NSSC20K0566 and NSF CAREER 1945310 grants.
PT acknowledges support from National Science Foundation (NSF) grants AST-1909933, AST-200849 and National Aeronautics and Space Administration (NASA) Astrophysics Theory Program (ATP) grant 80NSSC20K0502.

\appendix

\section*{Data availability}

The data that support the findings of this study are available from the corresponding author, upon reasonable request.

\bibliographystyle{mnras}
\bibliography{references} 

\begin{thebibliography}{}
\makeatletter
\relax
\def\mn@urlcharsother{\let\do\@makeother \do\$\do\&\do\#\do\^\do\_\do\%\do\~}
\def\mn@doi{\begingroup\mn@urlcharsother \@ifnextchar [ {\mn@doi@}
  {\mn@doi@[]}}
\def\mn@doi@[#1]#2{\def\@tempa{#1}\ifx\@tempa\@empty \href
  {http://dx.doi.org/#2} {doi:#2}\else \href {http://dx.doi.org/#2} {#1}\fi
  \endgroup}
\def\mn@eprint#1#2{\mn@eprint@#1:#2::\@nil}
\def\mn@eprint@arXiv#1{\href {http://arxiv.org/abs/#1} {{\tt arXiv:#1}}}
\def\mn@eprint@dblp#1{\href {http://dblp.uni-trier.de/rec/bibtex/#1.xml}
  {dblp:#1}}
\def\mn@eprint@#1:#2:#3:#4\@nil{\def\@tempa {#1}\def\@tempb {#2}\def\@tempc
  {#3}\ifx \@tempc \@empty \let \@tempc \@tempb \let \@tempb \@tempa \fi \ifx
  \@tempb \@empty \def\@tempb {arXiv}\fi \@ifundefined
  {mn@eprint@\@tempb}{\@tempb:\@tempc}{\expandafter \expandafter \csname
  mn@eprint@\@tempb\endcsname \expandafter{\@tempc}}}

\bibitem[\protect\citeauthoryear{{Adamo}, {Kruijssen}, {Bastian}, {Silva-Villa}
   \& {Ryon}}{{Adamo} et~al.}{2015}]{adamo_etal15}
{Adamo} A.,  {Kruijssen} J.~M.~D.,  {Bastian} N.,  {Silva-Villa} E.,   {Ryon}
  J.,  2015, \mn@doi [\mnras] {10.1093/mnras/stv1203}, \href
  {http://adsabs.harvard.edu/abs/2015MNRAS.452..246A} {452, 246}

\bibitem[\protect\citeauthoryear{{Adamo} et~al.,}{{Adamo}
  et~al.}{2020}]{adamo_etal20}
{Adamo} A.,  et~al., 2020, \mn@doi [\mnras] {10.1093/mnras/staa2380}, \href
  {https://ui.adsabs.harvard.edu/abs/2020MNRAS.499.3267A} {499, 3267}

\bibitem[\protect\citeauthoryear{{Ashman} \& {Zepf}}{{Ashman} \&
  {Zepf}}{1992}]{ashman_zepf92}
{Ashman} K.~M.,  {Zepf} S.~E.,  1992, \mn@doi [\apj] {10.1086/170850}, \href
  {http://adsabs.harvard.edu/abs/1992ApJ...384...50A} {384, 50}

\bibitem[\protect\citeauthoryear{{Barnes}}{{Barnes}}{1988}]{barnes88}
{Barnes} J.~E.,  1988, \mn@doi [\apj] {10.1086/166593}, \href
  {http://adsabs.harvard.edu/abs/1988ApJ...331..699B} {331, 699}

\bibitem[\protect\citeauthoryear{{Barnes}}{{Barnes}}{1992}]{barnes92}
{Barnes} J.~E.,  1992, \mn@doi [\apj] {10.1086/171522}, \href
  {https://ui.adsabs.harvard.edu/abs/1992ApJ...393..484B} {393, 484}

\bibitem[\protect\citeauthoryear{{Barnes}}{{Barnes}}{2004}]{barnes04}
{Barnes} J.~E.,  2004, \mn@doi [\mnras] {10.1111/j.1365-2966.2004.07725.x},
  \href {https://ui.adsabs.harvard.edu/abs/2004MNRAS.350..798B} {350, 798}

\bibitem[\protect\citeauthoryear{{Bastian}}{{Bastian}}{2008}]{bastian_08}
{Bastian} N.,  2008, \mn@doi [\mnras] {10.1111/j.1365-2966.2008.13775.x}, \href
  {http://adsabs.harvard.edu/abs/2008MNRAS.390..759B} {390, 759}

\bibitem[\protect\citeauthoryear{{Bastian}, {Schweizer}, {Goudfrooij}, {Larsen}
   \& {Kissler-Patig}}{{Bastian} et~al.}{2013}]{bastian_etal13}
{Bastian} N.,  {Schweizer} F.,  {Goudfrooij} P.,  {Larsen} S.~S.,
  {Kissler-Patig} M.,  2013, \mn@doi [\mnras] {10.1093/mnras/stt253}, \href
  {https://ui.adsabs.harvard.edu/abs/2013MNRAS.431.1252B} {431, 1252}

\bibitem[\protect\citeauthoryear{{Baumgardt} \& {Makino}}{{Baumgardt} \&
  {Makino}}{2003}]{baumgardt_makino03}
{Baumgardt} H.,  {Makino} J.,  2003, \mn@doi [\mnras]
  {10.1046/j.1365-8711.2003.06286.x}, \href
  {http://adsabs.harvard.edu/abs/2003MNRAS.340..227B} {340, 227}

\bibitem[\protect\citeauthoryear{{Bekki} \& {Chiba}}{{Bekki} \&
  {Chiba}}{2002}]{bekki_chiba02}
{Bekki} K.,  {Chiba} M.,  2002, \apj, 566, 245

\bibitem[\protect\citeauthoryear{{Bekki} \& {Couch}}{{Bekki} \&
  {Couch}}{2001}]{bekki_couch01}
{Bekki} K.,  {Couch} W.~J.,  2001, \mn@doi [\apjl] {10.1086/323139}, \href
  {https://ui.adsabs.harvard.edu/abs/2001ApJ...557L..19B} {557, L19}

\bibitem[\protect\citeauthoryear{{Benson}}{{Benson}}{2005}]{benson05}
{Benson} A.~J.,  2005, \mn@doi [\mnras] {10.1111/j.1365-2966.2005.08788.x},
  \href {https://ui.adsabs.harvard.edu/abs/2005MNRAS.358..551B} {358, 551}

\bibitem[\protect\citeauthoryear{{Bik}, {Lamers}, {Bastian}, {Panagia}  \&
  {Romaniello}}{{Bik} et~al.}{2003}]{bik_etal03}
{Bik} A.,  {Lamers} H.~J.~G.~L.~M.,  {Bastian} N.,  {Panagia} N.,
  {Romaniello} M.,  2003, \aap, 397, 473

\bibitem[\protect\citeauthoryear{{Bournaud}, {Jog}  \& {Combes}}{{Bournaud}
  et~al.}{2005}]{bournaud_etal05}
{Bournaud} F.,  {Jog} C.~J.,   {Combes} F.,  2005, \mn@doi [\aap]
  {10.1051/0004-6361:20042036}, \href
  {https://ui.adsabs.harvard.edu/abs/2005A&A...437...69B} {437, 69}

\bibitem[\protect\citeauthoryear{{Bournaud}, {Duc}  \& {Emsellem}}{{Bournaud}
  et~al.}{2008}]{bournaud_etal08}
{Bournaud} F.,  {Duc} P.~A.,   {Emsellem} E.,  2008, \mn@doi [\mnras]
  {10.1111/j.1745-3933.2008.00511.x}, \href
  {https://ui.adsabs.harvard.edu/abs/2008MNRAS.389L...8B} {389, L8}

\bibitem[\protect\citeauthoryear{{Bournaud} et~al.,}{{Bournaud}
  et~al.}{2011}]{bournaud_etal11}
{Bournaud} F.,  et~al., 2011, \mn@doi [\apj] {10.1088/0004-637X/730/1/4}, \href
  {https://ui.adsabs.harvard.edu/abs/2011ApJ...730....4B} {730, 4}

\bibitem[\protect\citeauthoryear{{Boutloukos} \& {Lamers}}{{Boutloukos} \&
  {Lamers}}{2003}]{boutloukos_lamers03}
{Boutloukos} S.~G.,  {Lamers} H.~J.~G.~L.~M.,  2003, \mn@doi [\mnras]
  {10.1046/j.1365-8711.2003.06083.x}, \href
  {https://ui.adsabs.harvard.edu/abs/2003MNRAS.338..717B} {338, 717}

\bibitem[\protect\citeauthoryear{{Braine} \& {Combes}}{{Braine} \&
  {Combes}}{1993}]{braine_combes93}
{Braine} J.,  {Combes} F.,  1993, \aap, \href
  {https://ui.adsabs.harvard.edu/abs/1993A&A...269....7B} {269, 7}

\bibitem[\protect\citeauthoryear{{Burkert}, {Naab}, {Johansson}  \&
  {Jesseit}}{{Burkert} et~al.}{2008}]{burkert_etal08}
{Burkert} A.,  {Naab} T.,  {Johansson} P.~H.,   {Jesseit} R.,  2008, \mn@doi
  [\apj] {10.1086/591632}, \href
  {https://ui.adsabs.harvard.edu/abs/2008ApJ...685..897B} {685, 897}

\bibitem[\protect\citeauthoryear{{Chabrier}}{{Chabrier}}{2003}]{chabrier03}
{Chabrier} G.,  2003, \mn@doi [\pasp] {10.1086/376392}, \href
  {http://adsabs.harvard.edu/abs/2003PASP..115..763C} {115, 763}

\bibitem[\protect\citeauthoryear{{Choksi} \& {Gnedin}}{{Choksi} \&
  {Gnedin}}{2019}]{choksi_gnedin19}
{Choksi} N.,  {Gnedin} O.~Y.,  2019, \mn@doi [\mnras] {10.1093/mnras/stz811},
  \href {https://ui.adsabs.harvard.edu/abs/2019MNRAS.486..331C} {486, 331}

\bibitem[\protect\citeauthoryear{{Choksi}, {Gnedin}  \& {Li}}{{Choksi}
  et~al.}{2018}]{choksi_etal18}
{Choksi} N.,  {Gnedin} O.~Y.,   {Li} H.,  2018, \mn@doi [\mnras]
  {10.1093/mnras/sty1952}, \href
  {http://adsabs.harvard.edu/abs/2018MNRAS.480.2343C} {480, 2343}

\bibitem[\protect\citeauthoryear{{Conselice}, {Chapman}  \&
  {Windhorst}}{{Conselice} et~al.}{2003}]{conselice_etal03}
{Conselice} C.~J.,  {Chapman} S.~C.,   {Windhorst} R.~A.,  2003, \mn@doi
  [\apjl] {10.1086/379109}, \href
  {https://ui.adsabs.harvard.edu/abs/2003ApJ...596L...5C} {596, L5}

\bibitem[\protect\citeauthoryear{{Cox}, {Jonsson}, {Primack}  \&
  {Somerville}}{{Cox} et~al.}{2006}]{cox_etal06}
{Cox} T.~J.,  {Jonsson} P.,  {Primack} J.~R.,   {Somerville} R.~S.,  2006,
  \mn@doi [\mnras] {10.1111/j.1365-2966.2006.11107.x}, \href
  {https://ui.adsabs.harvard.edu/abs/2006MNRAS.373.1013C} {373, 1013}

\bibitem[\protect\citeauthoryear{{Cox}, {Jonsson}, {Somerville}, {Primack}  \&
  {Dekel}}{{Cox} et~al.}{2008}]{cox_etal08}
{Cox} T.~J.,  {Jonsson} P.,  {Somerville} R.~S.,  {Primack} J.~R.,   {Dekel}
  A.,  2008, \mn@doi [\mnras] {10.1111/j.1365-2966.2007.12730.x}, \href
  {https://ui.adsabs.harvard.edu/abs/2008MNRAS.384..386C} {384, 386}

\bibitem[\protect\citeauthoryear{{Davies} et~al.,}{{Davies}
  et~al.}{2015}]{davies_etal15}
{Davies} L.~J.~M.,  et~al., 2015, \mn@doi [\mnras] {10.1093/mnras/stv1241},
  \href {https://ui.adsabs.harvard.edu/abs/2015MNRAS.452..616D} {452, 616}

\bibitem[\protect\citeauthoryear{{Di Matteo}, {Combes}, {Melchior}  \&
  {Semelin}}{{Di Matteo} et~al.}{2007}]{di-matteo_etal07}
{Di Matteo} P.,  {Combes} F.,  {Melchior} A.~L.,   {Semelin} B.,  2007, \mn@doi
  [\aap] {10.1051/0004-6361:20066959}, \href
  {https://ui.adsabs.harvard.edu/abs/2007A&A...468...61D} {468, 61}

\bibitem[\protect\citeauthoryear{{Ellison}, {Patton}, {Simard}  \&
  {McConnachie}}{{Ellison} et~al.}{2008}]{ellison_etal08}
{Ellison} S.~L.,  {Patton} D.~R.,  {Simard} L.,   {McConnachie} A.~W.,  2008,
  \mn@doi [\aj] {10.1088/0004-6256/135/5/1877}, \href
  {https://ui.adsabs.harvard.edu/abs/2008AJ....135.1877E} {135, 1877}

\bibitem[\protect\citeauthoryear{{Ellison}, {Catinella}  \&
  {Cortese}}{{Ellison} et~al.}{2018}]{ellison_etal18}
{Ellison} S.~L.,  {Catinella} B.,   {Cortese} L.,  2018, \mn@doi [\mnras]
  {10.1093/mnras/sty1247}, \href
  {https://ui.adsabs.harvard.edu/abs/2018MNRAS.478.3447E} {478, 3447}

\bibitem[\protect\citeauthoryear{{Forbes} \& {Hau}}{{Forbes} \&
  {Hau}}{2000}]{forbes_hau00}
{Forbes} D.~A.,  {Hau} G.~K.~T.,  2000, \mn@doi [\mnras]
  {10.1046/j.1365-8711.2000.03175.x}, \href
  {https://ui.adsabs.harvard.edu/abs/2000MNRAS.312..703F} {312, 703}

\bibitem[\protect\citeauthoryear{{Gieles} \& {Baumgardt}}{{Gieles} \&
  {Baumgardt}}{2008}]{gieles_baumgardt08}
{Gieles} M.,  {Baumgardt} H.,  2008, \mnras, \href
  {http://adsabs.harvard.edu/abs/2008MNRAS.389L..28G} {389, L28}

\bibitem[\protect\citeauthoryear{{Gieles}, {Larsen}, {Bastian}  \&
  {Stein}}{{Gieles} et~al.}{2006}]{gieles_etal06}
{Gieles} M.,  {Larsen} S.~S.,  {Bastian} N.,   {Stein} I.~T.,  2006, \mn@doi
  [\aap] {10.1051/0004-6361:20053589}, \href
  {http://adsabs.harvard.edu/abs/2006A%26A...450..129G} {450, 129}

\bibitem[\protect\citeauthoryear{{Harris}, {Whitmore}, {Karakla}, {Oko{\'n}},
  {Baum}, {Hanes}  \& {Kavelaars}}{{Harris} et~al.}{2006}]{harris_etal06}
{Harris} W.~E.,  {Whitmore} B.~C.,  {Karakla} D.,  {Oko{\'n}} W.,  {Baum}
  W.~A.,  {Hanes} D.~A.,   {Kavelaars} J.~J.,  2006, \mn@doi [\apj]
  {10.1086/498058}, \href {http://adsabs.harvard.edu/abs/2006ApJ...636...90H}
  {636, 90}

\bibitem[\protect\citeauthoryear{{Harris}, {Ciccone}, {Eadie}, {Gnedin},
  {Geisler}, {Rothberg}  \& {Bailin}}{{Harris} et~al.}{2017}]{harris_etal17b}
{Harris} W.~E.,  {Ciccone} S.~M.,  {Eadie} G.~M.,  {Gnedin} O.~Y.,  {Geisler}
  D.,  {Rothberg} B.,   {Bailin} J.,  2017, \mn@doi [\apj]
  {10.3847/1538-4357/835/1/101}, \href
  {http://adsabs.harvard.edu/abs/2017ApJ...835..101H} {835, 101}

\bibitem[\protect\citeauthoryear{{Hayward} et~al.,}{{Hayward}
  et~al.}{2014}]{hayward_etal14}
{Hayward} C.~C.,  et~al., 2014, \mn@doi [\mnras] {10.1093/mnras/stu1843}, \href
  {https://ui.adsabs.harvard.edu/abs/2014MNRAS.445.1598H} {445, 1598}

\bibitem[\protect\citeauthoryear{{Hernquist}}{{Hernquist}}{1992}]{hernquist92}
{Hernquist} L.,  1992, \mn@doi [\apj] {10.1086/172009}, \href
  {https://ui.adsabs.harvard.edu/abs/1992ApJ...400..460H} {400, 460}

\bibitem[\protect\citeauthoryear{{Hernquist}}{{Hernquist}}{1993}]{hernquist93}
{Hernquist} L.,  1993, \mn@doi [\apj] {10.1086/172686}, \href
  {https://ui.adsabs.harvard.edu/abs/1993ApJ...409..548H} {409, 548}

\bibitem[\protect\citeauthoryear{{Holmberg}}{{Holmberg}}{1941}]{holmberg41}
{Holmberg} E.,  1941, \mn@doi [\apj] {10.1086/144344}, \href
  {https://ui.adsabs.harvard.edu/abs/1941ApJ....94..385H} {94, 385}

\bibitem[\protect\citeauthoryear{{Holtzman} et~al.,}{{Holtzman}
  et~al.}{1992}]{holtzman_etal92}
{Holtzman} J.~A.,  et~al., 1992, \aj, 103, 691

\bibitem[\protect\citeauthoryear{{Holtzman} et~al.,}{{Holtzman}
  et~al.}{1996}]{holtzman_etal96}
{Holtzman} J.~A.,  et~al., 1996, \aj, 112, 416

\bibitem[\protect\citeauthoryear{{Hopkins}, {Cox}, {Hernquist}, {Narayanan},
  {Hayward}  \& {Murray}}{{Hopkins} et~al.}{2013}]{hopkins_etal13b}
{Hopkins} P.~F.,  {Cox} T.~J.,  {Hernquist} L.,  {Narayanan} D.,  {Hayward}
  C.~C.,   {Murray} N.,  2013, \mn@doi [\mnras] {10.1093/mnras/stt017}, \href
  {https://ui.adsabs.harvard.edu/abs/2013MNRAS.430.1901H} {430, 1901}

\bibitem[\protect\citeauthoryear{{Jesseit}, {Cappellari}, {Naab}, {Emsellem}
  \& {Burkert}}{{Jesseit} et~al.}{2009}]{jesseit_etal09}
{Jesseit} R.,  {Cappellari} M.,  {Naab} T.,  {Emsellem} E.,   {Burkert} A.,
  2009, \mn@doi [\mnras] {10.1111/j.1365-2966.2009.14984.x}, \href
  {https://ui.adsabs.harvard.edu/abs/2009MNRAS.397.1202J} {397, 1202}

\bibitem[\protect\citeauthoryear{{Johnson} et~al.,}{{Johnson}
  et~al.}{2017}]{johnson_etal17}
{Johnson} L.~C.,  et~al., 2017, \mn@doi [\apj] {10.3847/1538-4357/aa6a1f},
  \href {http://adsabs.harvard.edu/abs/2017ApJ...839...78J} {839, 78}

\bibitem[\protect\citeauthoryear{{Keller}, {Kruijssen}, {Pfeffer},
  {Reina-Campos}, {Bastian}, {Trujillo-Gomez}, {Hughes}  \& {Crain}}{{Keller}
  et~al.}{2020}]{keller_etal20}
{Keller} B.~W.,  {Kruijssen} J.~M.~D.,  {Pfeffer} J.,  {Reina-Campos} M.,
  {Bastian} N.,  {Trujillo-Gomez} S.,  {Hughes} M.~E.,   {Crain} R.~A.,  2020,
  \mn@doi [\mnras] {10.1093/mnras/staa1439}, \href
  {https://ui.adsabs.harvard.edu/abs/2020MNRAS.495.4248K} {495, 4248}

\bibitem[\protect\citeauthoryear{{Khochfar} \& {Burkert}}{{Khochfar} \&
  {Burkert}}{2006}]{khochfar_burkert06}
{Khochfar} S.,  {Burkert} A.,  2006, \mn@doi [\aap]
  {10.1051/0004-6361:20053241}, \href
  {https://ui.adsabs.harvard.edu/abs/2006A&A...445..403K} {445, 403}

\bibitem[\protect\citeauthoryear{{Kim} et~al.,}{{Kim}
  et~al.}{2018}]{kim_etal18_fire}
{Kim} J.-h.,  et~al., 2018, \mn@doi [\mnras] {10.1093/mnras/stx2994}, \href
  {https://ui.adsabs.harvard.edu/abs/2018MNRAS.474.4232K} {474, 4232}

\bibitem[\protect\citeauthoryear{{Knapen}, {Cisternas}  \&
  {Querejeta}}{{Knapen} et~al.}{2015}]{knapen_etal15}
{Knapen} J.~H.,  {Cisternas} M.,   {Querejeta} M.,  2015, \mn@doi [\mnras]
  {10.1093/mnras/stv2135}, \href
  {https://ui.adsabs.harvard.edu/abs/2015MNRAS.454.1742K} {454, 1742}

\bibitem[\protect\citeauthoryear{{Kravtsov} \& {Gnedin}}{{Kravtsov} \&
  {Gnedin}}{2005}]{kravtsov_gnedin05}
{Kravtsov} A.~V.,  {Gnedin} O.~Y.,  2005, \apj, 623, 650

\bibitem[\protect\citeauthoryear{{Kruijssen}}{{Kruijssen}}{2012}]{kruijssen12}
{Kruijssen} J.~M.~D.,  2012, \mn@doi [\mnras]
  {10.1111/j.1365-2966.2012.21923.x}, \href
  {http://adsabs.harvard.edu/abs/2012MNRAS.426.3008K} {426, 3008}

\bibitem[\protect\citeauthoryear{{Kruijssen}}{{Kruijssen}}{2015}]{kruijssen15}
{Kruijssen} J.~M.~D.,  2015, \mn@doi [\mnras] {10.1093/mnras/stv2026}, \href
  {http://adsabs.harvard.edu/abs/2015MNRAS.454.1658K} {454, 1658}

\bibitem[\protect\citeauthoryear{{Kruijssen}, {Pelupessy}, {Lamers}, {Portegies
  Zwart}  \& {Icke}}{{Kruijssen} et~al.}{2011}]{kruijssen_etal11}
{Kruijssen} J.~M.~D.,  {Pelupessy} F.~I.,  {Lamers} H. J.~G.~L.~M.,  {Portegies
  Zwart} S.~F.,   {Icke} V.,  2011, \mn@doi [\mnras]
  {10.1111/j.1365-2966.2011.18467.x}, \href
  {https://ui.adsabs.harvard.edu/abs/2011MNRAS.414.1339K} {414, 1339}

\bibitem[\protect\citeauthoryear{{Kruijssen}, {Pelupessy}, {Lamers}, {Portegies
  Zwart}, {Bastian}  \& {Icke}}{{Kruijssen} et~al.}{2012}]{kruijssen_etal12}
{Kruijssen} J.~M.~D.,  {Pelupessy} F.~I.,  {Lamers} H.~J.~G.~L.~M.,  {Portegies
  Zwart} S.~F.,  {Bastian} N.,   {Icke} V.,  2012, \mn@doi [\mnras]
  {10.1111/j.1365-2966.2012.20322.x}, \href
  {http://adsabs.harvard.edu/abs/2012MNRAS.421.1927K} {421, 1927}

\bibitem[\protect\citeauthoryear{{Kruijssen} et~al.,}{{Kruijssen}
  et~al.}{2020}]{kruijssen_etal20}
{Kruijssen} J.~M.~D.,  et~al., 2020, \mn@doi [\mnras] {10.1093/mnras/staa2452},
  \href {https://ui.adsabs.harvard.edu/abs/2020MNRAS.498.2472K} {498, 2472}

\bibitem[\protect\citeauthoryear{{Lacey} \& {Cole}}{{Lacey} \&
  {Cole}}{1993}]{lacey_cole93}
{Lacey} C.,  {Cole} S.,  1993, \mnras, \href
  {http://adsabs.harvard.edu/abs/1993MNRAS.262..627L} {262, 627}

\bibitem[\protect\citeauthoryear{{Lada} \& {Lada}}{{Lada} \&
  {Lada}}{2003}]{lada_lada03}
{Lada} C.~J.,  {Lada} E.~A.,  2003, \araa, 41, 57

\bibitem[\protect\citeauthoryear{{Lah{\'e}n}, {Naab}, {Johansson}, {Elmegreen},
  {Hu}, {Walch}, {Steinwandel}  \& {Moster}}{{Lah{\'e}n}
  et~al.}{2020}]{lahen_etal20}
{Lah{\'e}n} N.,  {Naab} T.,  {Johansson} P.~H.,  {Elmegreen} B.,  {Hu} C.-Y.,
  {Walch} S.,  {Steinwandel} U.~P.,   {Moster} B.~P.,  2020, \mn@doi [\apj]
  {10.3847/1538-4357/ab7190}, \href
  {https://ui.adsabs.harvard.edu/abs/2020ApJ...891....2L} {891, 2}

\bibitem[\protect\citeauthoryear{{Lamers}, {Gieles}  \& {Portegies
  Zwart}}{{Lamers} et~al.}{2005a}]{lamers_etal05}
{Lamers} H.~J.~G.~L.~M.,  {Gieles} M.,   {Portegies Zwart} S.~F.,  2005a,
  \mn@doi [\aap] {10.1051/0004-6361:20041476}, \href
  {http://adsabs.harvard.edu/abs/2005A%26A...429..173L} {429, 173}

\bibitem[\protect\citeauthoryear{{Lamers}, {Gieles}, {Bastian}, {Baumgardt},
  {Kharchenko}  \& {Portegies Zwart}}{{Lamers}
  et~al.}{2005b}]{lamers_etal05ana}
{Lamers} H.~J.~G.~L.~M.,  {Gieles} M.,  {Bastian} N.,  {Baumgardt} H.,
  {Kharchenko} N.~V.,   {Portegies Zwart} S.,  2005b, \mn@doi [\aap]
  {10.1051/0004-6361:20042241}, \href
  {https://ui.adsabs.harvard.edu/abs/2005A&A...441..117L} {441, 117}

\bibitem[\protect\citeauthoryear{{Larson} et~al.,}{{Larson}
  et~al.}{2016}]{larson_etal16}
{Larson} K.~L.,  et~al., 2016, \mn@doi [\apj] {10.3847/0004-637X/825/2/128},
  \href {https://ui.adsabs.harvard.edu/abs/2016ApJ...825..128L} {825, 128}

\bibitem[\protect\citeauthoryear{{Li} \& {Gnedin}}{{Li} \&
  {Gnedin}}{2014}]{li_gnedin14}
{Li} H.,  {Gnedin} O.~Y.,  2014, \mn@doi [\apj] {10.1088/0004-637X/796/1/10},
  \href {http://adsabs.harvard.edu/abs/2014ApJ...796...10L} {796, 10}

\bibitem[\protect\citeauthoryear{{Li} \& {Gnedin}}{{Li} \&
  {Gnedin}}{2019}]{li_gnedin19}
{Li} H.,  {Gnedin} O.~Y.,  2019, \mn@doi [\mnras] {10.1093/mnras/stz1114},
  \href {https://ui.adsabs.harvard.edu/abs/2019MNRAS.486.4030L} {486, 4030}

\bibitem[\protect\citeauthoryear{{Li}, {Mac Low}  \& {Klessen}}{{Li}
  et~al.}{2004}]{li_etal04}
{Li} Y.,  {Mac Low} M.,   {Klessen} R.~S.,  2004, \apjl, 614, L29

\bibitem[\protect\citeauthoryear{{Li}, {Gnedin}, {Gnedin}, {Meng}, {Semenov}
  \& {Kravtsov}}{{Li} et~al.}{2017}]{li_etal17}
{Li} H.,  {Gnedin} O.~Y.,  {Gnedin} N.~Y.,  {Meng} X.,  {Semenov} V.~A.,
  {Kravtsov} A.~V.,  2017, \mn@doi [\apj] {10.3847/1538-4357/834/1/69}, \href
  {http://adsabs.harvard.edu/abs/2017ApJ...834...69L} {834, 69}

\bibitem[\protect\citeauthoryear{{Li}, {Gnedin}  \& {Gnedin}}{{Li}
  et~al.}{2018}]{li_etal18}
{Li} H.,  {Gnedin} O.~Y.,   {Gnedin} N.~Y.,  2018, \mn@doi [\apj]
  {10.3847/1538-4357/aac9b8}, \href
  {http://adsabs.harvard.edu/abs/2018ApJ...861..107L} {861, 107}

\bibitem[\protect\citeauthoryear{{Li}, {Vogelsberger}, {Marinacci}  \&
  {Gnedin}}{{Li} et~al.}{2019}]{li_etal19}
{Li} H.,  {Vogelsberger} M.,  {Marinacci} F.,   {Gnedin} O.~Y.,  2019, \mn@doi
  [\mnras] {10.1093/mnras/stz1271}, \href
  {https://ui.adsabs.harvard.edu/abs/2019MNRAS.487..364L} {487, 364}

\bibitem[\protect\citeauthoryear{{Li}, {Vogelsberger}, {Marinacci}, {Sales}  \&
  {Torrey}}{{Li} et~al.}{2020}]{li_etal20}
{Li} H.,  {Vogelsberger} M.,  {Marinacci} F.,  {Sales} L.~V.,   {Torrey} P.,
  2020, \mn@doi [\mnras] {10.1093/mnras/staa3122}, \href
  {https://ui.adsabs.harvard.edu/abs/2020MNRAS.499.5862L} {499, 5862}

\bibitem[\protect\citeauthoryear{{Linden} et~al.,}{{Linden}
  et~al.}{2021}]{linden_etal21}
{Linden} S.~T.,  et~al., 2021, \mn@doi [\apj] {10.3847/1538-4357/ac2892}, \href
  {https://ui.adsabs.harvard.edu/abs/2021ApJ...923..278L} {923, 278}

\bibitem[\protect\citeauthoryear{{Lotz} et~al.,}{{Lotz}
  et~al.}{2008}]{lotz_etal08}
{Lotz} J.~M.,  et~al., 2008, \mn@doi [\apj] {10.1086/523659}, \href
  {https://ui.adsabs.harvard.edu/abs/2008ApJ...672..177L} {672, 177}

\bibitem[\protect\citeauthoryear{{Lotz}, {Jonsson}, {Cox}  \& {Primack}}{{Lotz}
  et~al.}{2010a}]{lotz_etal10mass}
{Lotz} J.~M.,  {Jonsson} P.,  {Cox} T.~J.,   {Primack} J.~R.,  2010a, \mn@doi
  [\mnras] {10.1111/j.1365-2966.2010.16268.x}, \href
  {https://ui.adsabs.harvard.edu/abs/2010MNRAS.404..575L} {404, 575}

\bibitem[\protect\citeauthoryear{{Lotz}, {Jonsson}, {Cox}  \& {Primack}}{{Lotz}
  et~al.}{2010b}]{lotz_etal10gas}
{Lotz} J.~M.,  {Jonsson} P.,  {Cox} T.~J.,   {Primack} J.~R.,  2010b, \mn@doi
  [\mnras] {10.1111/j.1365-2966.2010.16269.x}, \href
  {https://ui.adsabs.harvard.edu/abs/2010MNRAS.404..590L} {404, 590}

\bibitem[\protect\citeauthoryear{{Ma} et~al.,}{{Ma} et~al.}{2020}]{ma_etal20}
{Ma} X.,  et~al., 2020, \mn@doi [\mnras] {10.1093/mnras/staa527}, \href
  {https://ui.adsabs.harvard.edu/abs/2020MNRAS.493.4315M} {493, 4315}

\bibitem[\protect\citeauthoryear{{Maji}, {Zhu}, {Li}, {Charlton}, {Hernquist}
  \& {Knebe}}{{Maji} et~al.}{2017}]{maji_etal17}
{Maji} M.,  {Zhu} Q.,  {Li} Y.,  {Charlton} J.,  {Hernquist} L.,   {Knebe} A.,
  2017, \mn@doi [\apj] {10.3847/1538-4357/aa7aa1}, \href
  {https://ui.adsabs.harvard.edu/abs/2017ApJ...844..108M} {844, 108}

\bibitem[\protect\citeauthoryear{{Marinacci}, {Sales}, {Vogelsberger}, {Torrey}
   \& {Springel}}{{Marinacci} et~al.}{2019}]{marinacci_etal19}
{Marinacci} F.,  {Sales} L.~V.,  {Vogelsberger} M.,  {Torrey} P.,   {Springel}
  V.,  2019, \mn@doi [\mnras] {10.1093/mnras/stz2391}, \href
  {https://ui.adsabs.harvard.edu/abs/2019MNRAS.489.4233M} {489, 4233}

\bibitem[\protect\citeauthoryear{{Matsui} et~al.,}{{Matsui}
  et~al.}{2012}]{matsui_etal12}
{Matsui} H.,  et~al., 2012, \mn@doi [\apj] {10.1088/0004-637X/746/1/26}, \href
  {https://ui.adsabs.harvard.edu/abs/2012ApJ...746...26M} {746, 26}

\bibitem[\protect\citeauthoryear{{McLaughlin} \& {van der Marel}}{{McLaughlin}
  \& {van der Marel}}{2005}]{mclaughlin_vandermarel05}
{McLaughlin} D.~E.,  {van der Marel} R.~P.,  2005, \mn@doi [\apjs]
  {10.1086/497429}, \href {http://adsabs.harvard.edu/abs/2005ApJS..161..304M}
  {161, 304}

\bibitem[\protect\citeauthoryear{{Melnick} \& {Mirabel}}{{Melnick} \&
  {Mirabel}}{1990}]{melnick_mirabel90}
{Melnick} J.,  {Mirabel} I.~F.,  1990, \aap, \href
  {https://ui.adsabs.harvard.edu/abs/1990A&A...231L..19M} {231, L19}

\bibitem[\protect\citeauthoryear{{Miah}, {Sharples}  \& {Cho}}{{Miah}
  et~al.}{2015}]{miah_etal15}
{Miah} J.~A.,  {Sharples} R.~M.,   {Cho} J.,  2015, \mn@doi [\mnras]
  {10.1093/mnras/stu2735}, \href
  {https://ui.adsabs.harvard.edu/abs/2015MNRAS.447.3639M} {447, 3639}

\bibitem[\protect\citeauthoryear{{Mihos} \& {Hernquist}}{{Mihos} \&
  {Hernquist}}{1994}]{mihos_hernquist94}
{Mihos} J.~C.,  {Hernquist} L.,  1994, \mn@doi [\apjl] {10.1086/187460}, \href
  {https://ui.adsabs.harvard.edu/abs/1994ApJ...431L...9M} {431, L9}

\bibitem[\protect\citeauthoryear{{Mihos} \& {Hernquist}}{{Mihos} \&
  {Hernquist}}{1996}]{mihos_hernquist96}
{Mihos} J.~C.,  {Hernquist} L.,  1996, \mn@doi [\apj] {10.1086/177353}, \href
  {https://ui.adsabs.harvard.edu/abs/1996ApJ...464..641M} {464, 641}

\bibitem[\protect\citeauthoryear{{Moreno} et~al.,}{{Moreno}
  et~al.}{2019}]{moreno_etal19}
{Moreno} J.,  et~al., 2019, \mn@doi [\mnras] {10.1093/mnras/stz417}, \href
  {https://ui.adsabs.harvard.edu/abs/2019MNRAS.485.1320M} {485, 1320}

\bibitem[\protect\citeauthoryear{{Moreno} et~al.,}{{Moreno}
  et~al.}{2021}]{moreno_etal21}
{Moreno} J.,  et~al., 2021, \mn@doi [\mnras] {10.1093/mnras/staa2952}, \href
  {https://ui.adsabs.harvard.edu/abs/2021MNRAS.503.3113M} {503, 3113}

\bibitem[\protect\citeauthoryear{{Mortlock} et~al.,}{{Mortlock}
  et~al.}{2013}]{mortlock_etal13}
{Mortlock} A.,  et~al., 2013, \mn@doi [\mnras] {10.1093/mnras/stt793}, \href
  {https://ui.adsabs.harvard.edu/abs/2013MNRAS.433.1185M} {433, 1185}

\bibitem[\protect\citeauthoryear{{Muratov} \& {Gnedin}}{{Muratov} \&
  {Gnedin}}{2010}]{muratov_gnedin10}
{Muratov} A.~L.,  {Gnedin} O.~Y.,  2010, \mn@doi [\apj]
  {10.1088/0004-637X/718/2/1266}, \href
  {http://adsabs.harvard.edu/abs/2010ApJ...718.1266M} {718, 1266}

\bibitem[\protect\citeauthoryear{{Naab} \& {Burkert}}{{Naab} \&
  {Burkert}}{2003}]{naab_burkert03}
{Naab} T.,  {Burkert} A.,  2003, \mn@doi [\apj] {10.1086/378581}, \href
  {https://ui.adsabs.harvard.edu/abs/2003ApJ...597..893N} {597, 893}

\bibitem[\protect\citeauthoryear{{Negroponte} \& {White}}{{Negroponte} \&
  {White}}{1983}]{negroponte_white83}
{Negroponte} J.,  {White} S.~D.~M.,  1983, \mn@doi [\mnras]
  {10.1093/mnras/205.4.1009}, \href
  {https://ui.adsabs.harvard.edu/abs/1983MNRAS.205.1009N} {205, 1009}

\bibitem[\protect\citeauthoryear{{O'Connell}, {Gallagher}  \&
  {Hunter}}{{O'Connell} et~al.}{1994}]{oconnell_etal94}
{O'Connell} R.~W.,  {Gallagher} John~S. I.,   {Hunter} D.~A.,  1994, \mn@doi
  [\apj] {10.1086/174625}, \href
  {https://ui.adsabs.harvard.edu/abs/1994ApJ...433...65O} {433, 65}

\bibitem[\protect\citeauthoryear{{Patton}, {Torrey}, {Ellison}, {Mendel}  \&
  {Scudder}}{{Patton} et~al.}{2013}]{patton_etal13}
{Patton} D.~R.,  {Torrey} P.,  {Ellison} S.~L.,  {Mendel} J.~T.,   {Scudder}
  J.~M.,  2013, \mn@doi [\mnras] {10.1093/mnrasl/slt058}, \href
  {https://ui.adsabs.harvard.edu/abs/2013MNRAS.433L..59P} {433, L59}

\bibitem[\protect\citeauthoryear{{Peng} et~al.,}{{Peng}
  et~al.}{2006}]{peng_etal06}
{Peng} E.~W.,  et~al., 2006, \mn@doi [\apj] {10.1086/498210}, \href
  {https://ui.adsabs.harvard.edu/abs/2006ApJ...639...95P} {639, 95}

\bibitem[\protect\citeauthoryear{{Pfeffer}, {Kruijssen}, {Crain}  \&
  {Bastian}}{{Pfeffer} et~al.}{2018}]{pfeffer_etal18}
{Pfeffer} J.,  {Kruijssen} J.~M.~D.,  {Crain} R.~A.,   {Bastian} N.,  2018,
  \mn@doi [\mnras] {10.1093/mnras/stx3124}, \href
  {http://adsabs.harvard.edu/abs/2018MNRAS.475.4309P} {475, 4309}

\bibitem[\protect\citeauthoryear{{Pfleiderer} \& {Siedentopf}}{{Pfleiderer} \&
  {Siedentopf}}{1961}]{pfleiderer_siedentopf61}
{Pfleiderer} J.,  {Siedentopf} H.,  1961, \zap, \href
  {https://ui.adsabs.harvard.edu/abs/1961ZA.....51..201P} {51, 201}

\bibitem[\protect\citeauthoryear{{Portegies Zwart}, {McMillan}  \&
  {Gieles}}{{Portegies Zwart} et~al.}{2010a}]{portegies_etal10}
{Portegies Zwart} S.~F.,  {McMillan} S.~L.~W.,   {Gieles} M.,  2010a, \mn@doi
  [\araa] {10.1146/annurev-astro-081309-130834}, \href
  {http://adsabs.harvard.edu/abs/2010ARA%26A..48..431P} {48, 431}

\bibitem[\protect\citeauthoryear{{Portegies Zwart}, {McMillan}  \&
  {Gieles}}{{Portegies Zwart} et~al.}{2010b}]{portegies_zwart_etal10}
{Portegies Zwart} S.~F.,  {McMillan} S.~L.~W.,   {Gieles} M.,  2010b, \mn@doi
  [\araa] {10.1146/annurev-astro-081309-130834}, \href
  {http://adsabs.harvard.edu/abs/2010ARA%26A..48..431P} {48, 431}

\bibitem[\protect\citeauthoryear{{Powell}, {Bournaud}, {Chapon}  \&
  {Teyssier}}{{Powell} et~al.}{2013}]{powell_etal13}
{Powell} L.~C.,  {Bournaud} F.,  {Chapon} D.,   {Teyssier} R.,  2013, \mn@doi
  [\mnras] {10.1093/mnras/stt1036}, \href
  {https://ui.adsabs.harvard.edu/abs/2013MNRAS.434.1028P} {434, 1028}

\bibitem[\protect\citeauthoryear{{Randriamanakoto}, {V{\"a}is{\"a}nen}, {Ryder}
   \& {Ranaivomanana}}{{Randriamanakoto} et~al.}{2019}]{randriamanakoto_etal19}
{Randriamanakoto} Z.,  {V{\"a}is{\"a}nen} P.,  {Ryder} S.~D.,   {Ranaivomanana}
  P.,  2019, \mn@doi [\mnras] {10.1093/mnras/sty2837}, \href
  {https://ui.adsabs.harvard.edu/abs/2019MNRAS.482.2530R} {482, 2530}

\bibitem[\protect\citeauthoryear{{Reina-Campos}, {Keller}, {Kruijssen},
  {Gensior}, {Trujillo-Gomez}, {Jeffreson}, {Pfeffer}  \&
  {Sills}}{{Reina-Campos} et~al.}{2022}]{reina-campos_etal22}
{Reina-Campos} M.,  {Keller} B.~W.,  {Kruijssen} J.~M.~D.,  {Gensior} J.,
  {Trujillo-Gomez} S.,  {Jeffreson} S. M.~R.,  {Pfeffer} J.~L.,   {Sills} A.,
  2022, arXiv e-prints, \href
  {https://ui.adsabs.harvard.edu/abs/2022arXiv220206961R} {p. arXiv:2202.06961}

\bibitem[\protect\citeauthoryear{{Reines}, {Johnson}  \& {Goss}}{{Reines}
  et~al.}{2008}]{reines_etal08}
{Reines} A.~E.,  {Johnson} K.~E.,   {Goss} W.~M.,  2008, \mn@doi [\aj]
  {10.1088/0004-6256/135/6/2222}, \href
  {http://adsabs.harvard.edu/abs/2008AJ....135.2222R} {135, 2222}

\bibitem[\protect\citeauthoryear{{Renaud} \& {Gieles}}{{Renaud} \&
  {Gieles}}{2013}]{renaud_gieles13}
{Renaud} F.,  {Gieles} M.,  2013, \mn@doi [\mnras] {10.1093/mnrasl/slt013},
  \href {https://ui.adsabs.harvard.edu/abs/2013MNRAS.431L..83R} {431, L83}

\bibitem[\protect\citeauthoryear{{Renaud}, {Gieles}  \& {Boily}}{{Renaud}
  et~al.}{2011}]{renaud_etal11}
{Renaud} F.,  {Gieles} M.,   {Boily} C.~M.,  2011, \mn@doi [\mnras]
  {10.1111/j.1365-2966.2011.19531.x}, \href
  {http://adsabs.harvard.edu/abs/2011MNRAS.418..759R} {418, 759}

\bibitem[\protect\citeauthoryear{{Renaud}, {Bournaud}  \& {Duc}}{{Renaud}
  et~al.}{2015}]{renaud_etal15}
{Renaud} F.,  {Bournaud} F.,   {Duc} P.-A.,  2015, \mn@doi [\mnras]
  {10.1093/mnras/stu2208}, \href
  {http://adsabs.harvard.edu/abs/2015MNRAS.446.2038R} {446, 2038}

\bibitem[\protect\citeauthoryear{{Renaud}, {Agertz}  \& {Gieles}}{{Renaud}
  et~al.}{2017}]{renaud_etal17}
{Renaud} F.,  {Agertz} O.,   {Gieles} M.,  2017, \mn@doi [\mnras]
  {10.1093/mnras/stw2969}, \href
  {http://adsabs.harvard.edu/abs/2017MNRAS.465.3622R} {465, 3622}

\bibitem[\protect\citeauthoryear{{Renaud}, {Bournaud}, {Agertz}, {Kraljic},
  {Schinnerer}, {Bolatto}, {Daddi}  \& {Hughes}}{{Renaud}
  et~al.}{2019}]{renaud_etal19}
{Renaud} F.,  {Bournaud} F.,  {Agertz} O.,  {Kraljic} K.,  {Schinnerer} E.,
  {Bolatto} A.,  {Daddi} E.,   {Hughes} A.,  2019, \mn@doi [\aap]
  {10.1051/0004-6361/201935222}, \href
  {https://ui.adsabs.harvard.edu/abs/2019A&A...625A..65R} {625, A65}

\bibitem[\protect\citeauthoryear{{Rieder}, {Ishiyama}, {Langelaan}, {Makino},
  {McMillan}  \& {Portegies Zwart}}{{Rieder} et~al.}{2013}]{rieder_etal13}
{Rieder} S.,  {Ishiyama} T.,  {Langelaan} P.,  {Makino} J.,  {McMillan}
  S.~L.~W.,   {Portegies Zwart} S.,  2013, \mn@doi [\mnras]
  {10.1093/mnras/stt1848}, \href
  {http://adsabs.harvard.edu/abs/2013MNRAS.436.3695R} {436, 3695}

\bibitem[\protect\citeauthoryear{{Robertson}, {Bullock}, {Cox}, {Di Matteo},
  {Hernquist}, {Springel}  \& {Yoshida}}{{Robertson}
  et~al.}{2006}]{robertson_etal06}
{Robertson} B.,  {Bullock} J.~S.,  {Cox} T.~J.,  {Di Matteo} T.,  {Hernquist}
  L.,  {Springel} V.,   {Yoshida} N.,  2006, \mn@doi [\apj] {10.1086/504412},
  \href {https://ui.adsabs.harvard.edu/abs/2006ApJ...645..986R} {645, 986}

\bibitem[\protect\citeauthoryear{{Rosolowsky}, {Pineda}, {Kauffmann}  \&
  {Goodman}}{{Rosolowsky} et~al.}{2008}]{rosolowsky_etal08}
{Rosolowsky} E.~W.,  {Pineda} J.~E.,  {Kauffmann} J.,   {Goodman} A.~A.,  2008,
  \mn@doi [\apj] {10.1086/587685}, \href
  {https://ui.adsabs.harvard.edu/abs/2008ApJ...679.1338R} {679, 1338}

\bibitem[\protect\citeauthoryear{{Saitoh}, {Daisaka}, {Kokubo}, {Makino},
  {Okamoto}, {Tomisaka}, {Wada}  \& {Yoshida}}{{Saitoh}
  et~al.}{2009}]{saitoh_etal09}
{Saitoh} T.~R.,  {Daisaka} H.,  {Kokubo} E.,  {Makino} J.,  {Okamoto} T.,
  {Tomisaka} K.,  {Wada} K.,   {Yoshida} N.,  2009, \mn@doi [\pasj]
  {10.1093/pasj/61.3.481}, \href
  {https://ui.adsabs.harvard.edu/abs/2009PASJ...61..481S} {61, 481}

\bibitem[\protect\citeauthoryear{{Sanders}, {Soifer}, {Elias}, {Madore},
  {Matthews}, {Neugebauer}  \& {Scoville}}{{Sanders}
  et~al.}{1988}]{sanders_etal88}
{Sanders} D.~B.,  {Soifer} B.~T.,  {Elias} J.~H.,  {Madore} B.~F.,  {Matthews}
  K.,  {Neugebauer} G.,   {Scoville} N.~Z.,  1988, \mn@doi [\apj]
  {10.1086/165983}, \href
  {https://ui.adsabs.harvard.edu/abs/1988ApJ...325...74S} {325, 74}

\bibitem[\protect\citeauthoryear{{Schweizer}}{{Schweizer}}{2004}]{schweizer04}
{Schweizer} F.,  2004, in ASP Conf. Ser. 322: The Formation and Evolution of
  Massive Young Star Clusters. p.~111

\bibitem[\protect\citeauthoryear{{Schweizer} \& {Seitzer}}{{Schweizer} \&
  {Seitzer}}{1998}]{schweizer_seitzer98}
{Schweizer} F.,  {Seitzer} P.,  1998, \mn@doi [\aj] {10.1086/300616}, \href
  {https://ui.adsabs.harvard.edu/abs/1998AJ....116.2206S} {116, 2206}

\bibitem[\protect\citeauthoryear{{Somerville} \& {Dav{\'e}}}{{Somerville} \&
  {Dav{\'e}}}{2015}]{somerville_dave15}
{Somerville} R.~S.,  {Dav{\'e}} R.,  2015, \mn@doi [\araa]
  {10.1146/annurev-astro-082812-140951}, \href
  {http://adsabs.harvard.edu/abs/2015ARA%26A..53...51S} {53, 51}

\bibitem[\protect\citeauthoryear{{Spitler} \& {Forbes}}{{Spitler} \&
  {Forbes}}{2009}]{spitler_forbes09}
{Spitler} L.~R.,  {Forbes} D.~A.,  2009, \mn@doi [\mnras]
  {10.1111/j.1745-3933.2008.00567.x}, \href
  {http://adsabs.harvard.edu/abs/2009MNRAS.392L...1S} {392, L1}

\bibitem[\protect\citeauthoryear{{Springel}}{{Springel}}{2010}]{springel10}
{Springel} V.,  2010, \mn@doi [\araa] {10.1146/annurev-astro-081309-130914},
  \href {http://adsabs.harvard.edu/abs/2010ARA%26A..48..391S} {48, 391}

\bibitem[\protect\citeauthoryear{{Springel}, {White}, {Tormen}  \&
  {Kauffmann}}{{Springel} et~al.}{2001}]{springel_etal01}
{Springel} V.,  {White} S. D.~M.,  {Tormen} G.,   {Kauffmann} G.,  2001,
  \mn@doi [\mnras] {10.1046/j.1365-8711.2001.04912.x}, \href
  {https://ui.adsabs.harvard.edu/abs/2001MNRAS.328..726S} {328, 726}

\bibitem[\protect\citeauthoryear{{Springel}, {Di Matteo}  \&
  {Hernquist}}{{Springel} et~al.}{2005}]{springel_etal05b}
{Springel} V.,  {Di Matteo} T.,   {Hernquist} L.,  2005, \mn@doi [\mnras]
  {10.1111/j.1365-2966.2005.09238.x}, \href
  {https://ui.adsabs.harvard.edu/abs/2005MNRAS.361..776S} {361, 776}

\bibitem[\protect\citeauthoryear{{Teyssier}, {Chapon}  \&
  {Bournaud}}{{Teyssier} et~al.}{2010}]{teyssier_etal10}
{Teyssier} R.,  {Chapon} D.,   {Bournaud} F.,  2010, \mn@doi [\apjl]
  {10.1088/2041-8205/720/2/L149}, \href
  {https://ui.adsabs.harvard.edu/abs/2010ApJ...720L.149T} {720, L149}

\bibitem[\protect\citeauthoryear{{Toomre}}{{Toomre}}{1977}]{toomre77}
{Toomre} A.,  1977, in {Tinsley} B.~M.,  {Larson} Richard B.~Gehret D.~C.,
  eds, Evolution of Galaxies and Stellar Populations. p.~401

\bibitem[\protect\citeauthoryear{{Toomre} \& {Toomre}}{{Toomre} \&
  {Toomre}}{1972}]{toomre_toomre72}
{Toomre} A.,  {Toomre} J.,  1972, \mn@doi [\apj] {10.1086/151823}, \href
  {http://adsabs.harvard.edu/abs/1972ApJ...178..623T} {178, 623}

\bibitem[\protect\citeauthoryear{{Tress}, {Smith}, {Sormani}, {Glover},
  {Klessen}, {Mac Low}  \& {Clark}}{{Tress} et~al.}{2020}]{tress_etal20}
{Tress} R.~G.,  {Smith} R.~J.,  {Sormani} M.~C.,  {Glover} S. C.~O.,  {Klessen}
  R.~S.,  {Mac Low} M.-M.,   {Clark} P.~C.,  2020, \mn@doi [\mnras]
  {10.1093/mnras/stz3600}, \href
  {https://ui.adsabs.harvard.edu/abs/2020MNRAS.492.2973T} {492, 2973}

\bibitem[\protect\citeauthoryear{{Ueda} et~al.,}{{Ueda}
  et~al.}{2014}]{ueda_etal14}
{Ueda} J.,  et~al., 2014, \mn@doi [\apjs] {10.1088/0067-0049/214/1/1}, \href
  {https://ui.adsabs.harvard.edu/abs/2014ApJS..214....1U} {214, 1}

\bibitem[\protect\citeauthoryear{{Usher} et~al.,}{{Usher}
  et~al.}{2012}]{usher_etal12}
{Usher} C.,  et~al., 2012, \mn@doi [\mnras] {10.1111/j.1365-2966.2012.21801.x},
  \href {http://adsabs.harvard.edu/abs/2012MNRAS.426.1475U} {426, 1475}

\bibitem[\protect\citeauthoryear{{Violino}, {Ellison}, {Sargent}, {Coppin},
  {Scudder}, {Mendel}  \& {Saintonge}}{{Violino} et~al.}{2018}]{violino_etal18}
{Violino} G.,  {Ellison} S.~L.,  {Sargent} M.,  {Coppin} K. E.~K.,  {Scudder}
  J.~M.,  {Mendel} T.~J.,   {Saintonge} A.,  2018, \mn@doi [\mnras]
  {10.1093/mnras/sty345}, \href
  {https://ui.adsabs.harvard.edu/abs/2018MNRAS.476.2591V} {476, 2591}

\bibitem[\protect\citeauthoryear{{Whitmore} \& {Schweizer}}{{Whitmore} \&
  {Schweizer}}{1995}]{whitmore_schweizer95}
{Whitmore} B.~C.,  {Schweizer} F.,  1995, \aj, 109, 960

\bibitem[\protect\citeauthoryear{{Whitmore}, {Zhang}, {Leitherer}, {Fall},
  {Schweizer}  \& {Miller}}{{Whitmore} et~al.}{1999}]{whitmore_etal99}
{Whitmore} B.~C.,  {Zhang} Q.,  {Leitherer} C.,  {Fall} S.~M.,  {Schweizer} F.,
    {Miller} B.~W.,  1999, \mn@doi [\aj] {10.1086/301041}, \href
  {http://adsabs.harvard.edu/abs/1999AJ....118.1551W} {118, 1551}

\bibitem[\protect\citeauthoryear{{Wong} et~al.,}{{Wong}
  et~al.}{2011}]{wong_etal11}
{Wong} K.~C.,  et~al., 2011, \mn@doi [\apj] {10.1088/0004-637X/728/2/119},
  \href {https://ui.adsabs.harvard.edu/abs/2011ApJ...728..119W} {728, 119}

\bibitem[\protect\citeauthoryear{{Zepf}, {Ashman}, {English}, {Freeman}  \&
  {Sharples}}{{Zepf} et~al.}{1999}]{zepf_etal99}
{Zepf} S.~E.,  {Ashman} K.~M.,  {English} J.,  {Freeman} K.~C.,   {Sharples}
  R.~M.,  1999, \aj, 118, 752

\bibitem[\protect\citeauthoryear{{Zhang} \& {Fall}}{{Zhang} \&
  {Fall}}{1999}]{zhang_fall99}
{Zhang} Q.,  {Fall} S.~M.,  1999, \apjl, 527, L81

\bibitem[\protect\citeauthoryear{{Zwicky}}{{Zwicky}}{1956}]{zwicky56}
{Zwicky} F.,  1956, Ergebnisse der exakten Naturwissenschaften, \href
  {https://ui.adsabs.harvard.edu/abs/1956ErNW...29..344Z} {29, 344}

\bibitem[\protect\citeauthoryear{{van den Bosch}, {Lewis}, {Lake}  \&
  {Stadel}}{{van den Bosch} et~al.}{1999}]{vandenbosch_etal99}
{van den Bosch} F.~C.,  {Lewis} G.~F.,  {Lake} G.,   {Stadel} J.,  1999,
  \mn@doi [\apj] {10.1086/307023}, \href
  {http://adsabs.harvard.edu/abs/1999ApJ...515...50V} {515, 50}

\makeatother
\end{thebibliography}

\appendix

\section{Numerical convergence on the mass function of YMCs}

\begin{figure}
\includegraphics[width=\columnwidth]{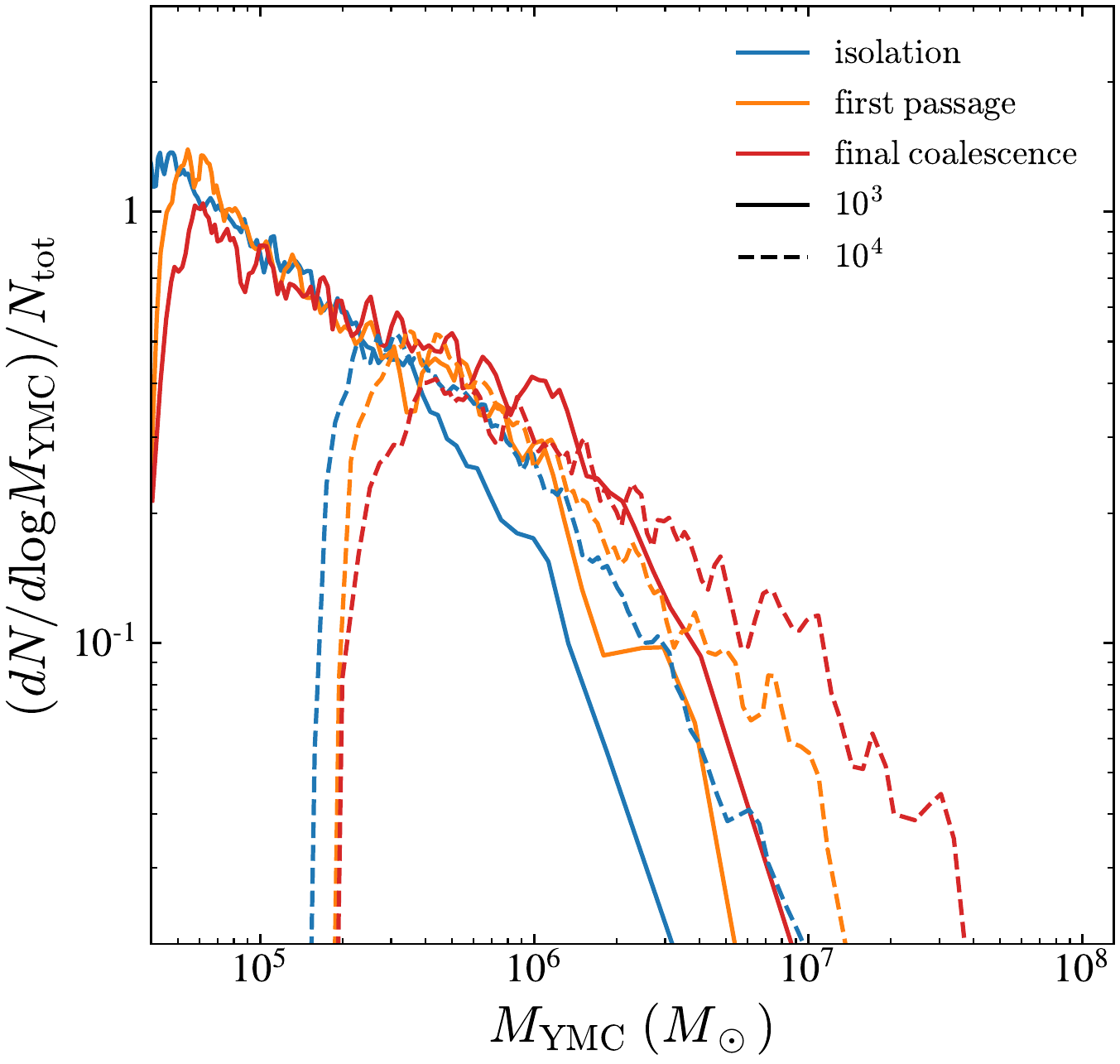}
\vspace{0mm}
\caption{Numerical convergence of the YMC mass functions for simulations with mass resolution of $10^3\Msun$ (solid) and $10^4\Msun$ (dashed) for three different evolutionary stages.}
  \label{fig:YMC-convergence}
\end{figure}

One of the main conclusions of this paper, as is shown in \autoref{sec:results-massfunction}, is that the mass function of YMCs is shallower and extends to higher masses during mergers. This result is derived from the merger simulation with fiducial mass resolution of $10^3\Msun$ (``1e3'' run). To test whether this is robust to numerical resolution, we perform another simulation with the same merger setup but different mass resolution of $10^4\Msun$ (``1e4'' run). Following the same YMC identification method as the fiducial run, we extract the YMC mass function from the 1e4 run in exactly the same three epochs. In \autoref{fig:YMC-convergence}, we show the comparison of the YMC mass functions in three different epochs for both the 1e3 and 1e4 runs. First, we find that, by design, the low-mass cutoff of the mass function is set by the numerical resolution and the SUBFIND minimum number of particles. For the 1e4 run, only YMCs with masses higher than $2-3\times10^5\Msun$ are resolved because of these limits, as well as artificial tidal disruptions.
The 1e4 run also shows a mass function extending to higher masses than the 1e3 run, probably because of the inability to resolve the turbulent cascade and fragmentation of GMCs in dense regions, especially at the center of the merger remnant.
Interestingly, despite the differences, the power-law slopes of the mass functions of the 1e4 runs are consistent with the 1e3 run for all three merger stages, with the same trend of shallower slope during mergers than those in isolation. Consequently, the overall cluster formation efficiency is significantly enhanced during mergers in the 1e4 run as well. These similarities suggest that the main conclusions in this paper are robust to numerical resolution.

It should be noted that we cannot perform higher resolution simulations than the 1e3 runs (e.g. mass resolution of 100 $\Msun$) because 1) such simulation is unfeasible because the computational cost would become prohibitively expensive to reach the final coalescence stage for a fair comparison 2) not suitable for the SMUGGLE model because each star particle in SMUGGLE represents a single stellar population that fully samples the IMF.

\section{Time evolution of the YMC populations in a mass-dependent tidal field}\label{sec:discussion-tides}

\begin{figure}
\includegraphics[width=\columnwidth]{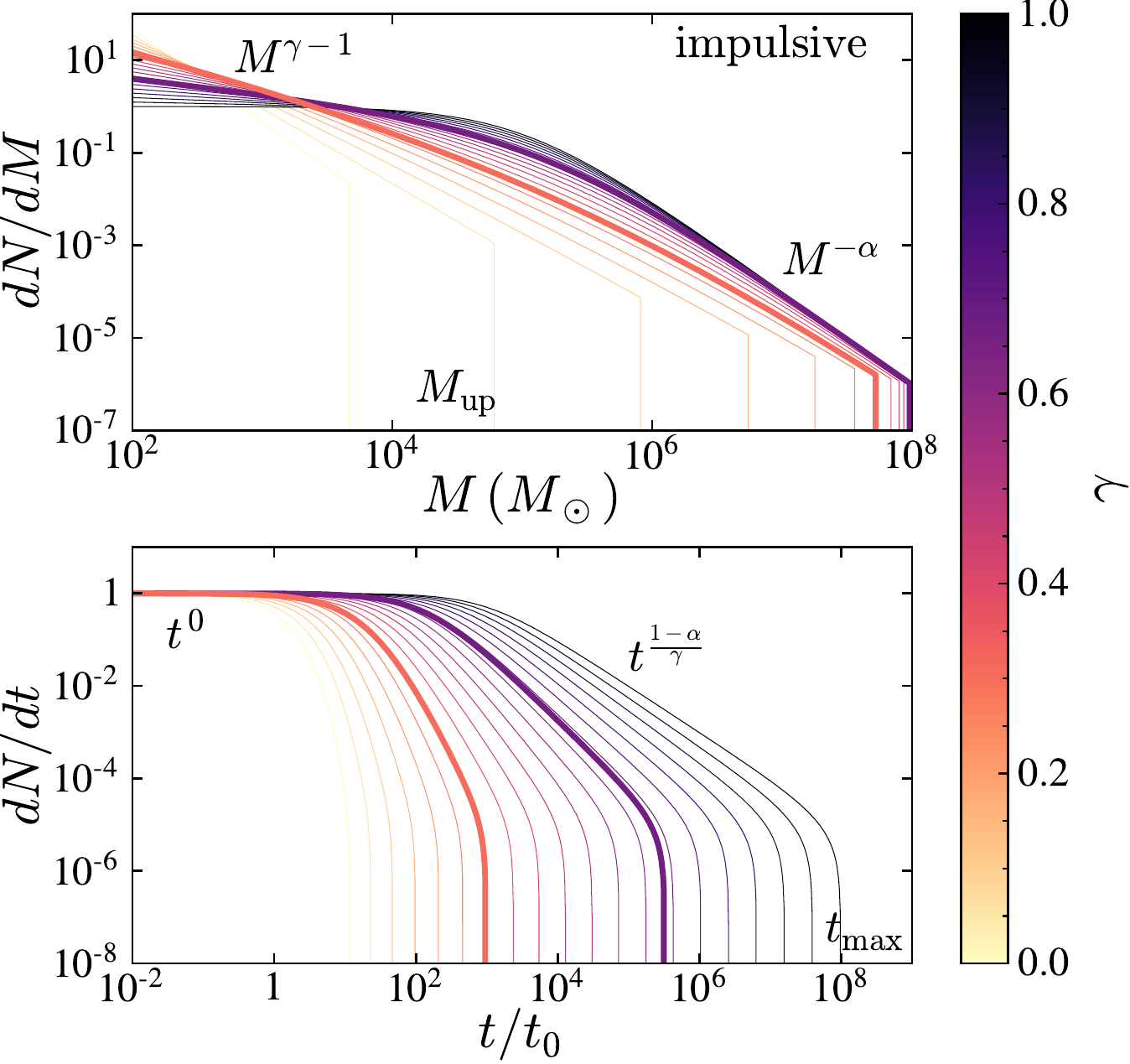}
\vspace{0mm}
\caption{Upper: Evolved YMC mass functions for different values of the mass-dependent tidal disruption timescale index $\gamma$, for the impulsive injection case. 
The normalization of the mass function is arbitrary, but in order to better show the behavior of the slope changes with $\gamma$, we enforce that $t_4 = 10^{4\gamma} t_0$ is a constant at $M=10^4\Msun$ for all lines. We also write down the asymptotic power-law slopes in both the low- and high- mass ends of the mass function together with the location of the upper limit, $M_{\rm up}$. Lower: The time evolution of the total number of clusters above a given minimum mass $\mmin$ for the impulsive injection case. All lines are normalized so that $dN/dt=1$ at $t=0$. For both panels, the color of the lines represent results with different values of $\gamma$. In particular, the case of $\gamma=2/3$ and $0.31$ are highlighted as thick lines, see \autoref{sec:discussion-tides} in detail. }
  \label{fig:analytical-YMC-impulsive}
\end{figure}

Stars are formed in clustered environments \citep[][]{lada_lada03}. Right after their formation, most clusters are disrupted and their member stars join the field. The disruption process strongly depends on the local environments around the clusters. Previous works have shown that young cluster populations in different galaxies experience drastically different tidal interactions \citep[e.g.][]{boutloukos_lamers03, lamers_etal05, randriamanakoto_etal19}.
In \autoref{sec:results-tides-mass}, we show that, even within a single galaxy, clusters with different initial masses experience systematically different tidal disruptions during the early stage of their evolution because they emerge from different overdensities in the hierarchical ISM. This mass-dependency inevitably leave its imprint on the time evolution of the total number and mass function of surviving clusters and can be tested by future observations of YMC populations.

Here we derive an analytical formalism for the time evolution of cluster populations from a general scaling between the tidal disruption timescale and cluster mass
\begin{equation}\label{eq:ttid}
    \ttid(M) = t_0 \left(\frac{M}{\Msun}\right)^\gamma
\end{equation}
with $\gamma>0$.
If, for simplicity, we ignore the mass loss from stellar evolution of individual stars, the mass loss of clusters only comes from tidal disruption that is characterized by tidal disruption timescale in \autoref{eq:ttid}:
\begin{equation}\label{eq:dmdt}
    \frac{dM}{dt} = \left(\frac{dM}{dt}\right)_{\rm tid} = - \frac{M}{\ttid} = 
    - \frac{M_\odot}{t_0}\left(\frac{M}{M_\odot}\right)^{1-\gamma}.
\end{equation}
For a cluster with an initial mass $M_i$, the time evolution of its mass can be obtained by integrating \autoref{eq:dmdt} over t:
\begin{equation}\label{eq:mtmi}
    \frac{M(t)}{M_i} = \left[1-\frac{\gamma t}{t_0}\left(\frac{M_\odot}{M_i}\right)^{\gamma} \right]^{1/\gamma}.
\end{equation}
This relation is only valid when $t<t_0(M_i/\Msun)^\gamma/\gamma$. Older clusters are completed disrupted. The derivation below largely follows \citet[][]{lamers_etal05ana}.

\subsubsection{Impulsive injection case}

First, we examine the time evolution of a single population of YMCs that are formed at the same epoch with a power-law CIMF, $dN/dM\propto M^{-\alpha}$. We call this the instantaneous injection case. The evolved mass function at an age $t$ can be derived from the conservation of the number of clusters: 
$dN(M,t) dM = dN(M_i,t) dM_i$. Therefore, the mass function at time $t$ is:
\begin{equation}\label{eq:dndmdt}
    \frac{dN_{\rm imp}(M,t)}{dM} \propto M^{-\alpha}\left[1+\frac{\gamma t}{t_0}\left(\frac{M}{\Msun}\right)^{-\gamma}\right]^\frac{1-\alpha-\gamma}{\gamma}.
\end{equation}
The upper panel of \autoref{fig:analytical-YMC-impulsive} shows the mass function of surviving clusters. 
For more massive (or younger) clusters when $\frac{\gamma t}{t_0}\left(\frac{M}{\Msun}\right)^{-\gamma}\ll1$, the slope of the mass function is $\alpha$, the same as the CIMF. On the other hand, for less massive (or older) clusters, the slope of the mass function becomes $\gamma-1$ and completely loses its memory of the initial value. Keep in mind that, if there exists a maximum cluster mass, $\mmax$, for the CIMF, \autoref{eq:dndmdt} is only valid for cluster masses below $M_{\rm up}=\mmax\left[1-\frac{\gamma t}{t_0}\left(\frac{\mmax}{\Msun}\right)^{-\gamma}\right]^{1/\gamma}$. This upper limit can also be clearly seen in the figure.

The time evolution of the total number of surviving clusters at an age $t$ can be obtained by integrating \autoref{eq:dndmdt} over a mass range between $\mmin$ and $M_{\rm up}$ (valid only when $M_{\rm up}>\mmin$), where $\mmin$ can be interpreted as the observational limit:
\begin{multline}\label{eq:nt-impulsive}
N_{\rm imp}(t) = \int_{\mmin}^{M_{\rm up}} \frac{dN_{\rm imp}(M,t)}{dM}dM \propto \\
\frac{1}{1-\alpha}\left\{\left(\frac{\mmax}{\Msun}\right)^{1-\alpha}-\left[\left(\frac{\mmin}{\Msun}\right)^\gamma+\frac{\gamma t}{t_0}\right]^{(1-\alpha)/\gamma}\right\}.
\end{multline}
The lower panel of \autoref{fig:analytical-YMC-impulsive} summarizes several key features of the evolution in different regimes with different values of $\gamma$.
In the very early stage when $\frac{\gamma t}{t_0}\ll \left(\frac{\mmin}{\Msun}\right)^\gamma$, 
the number of clusters is nearly constant $N_{\rm imp}(t)\sim \frac{1}{1-\alpha}\left[\left(\frac{\mmax}{\Msun}\right)^{1-\alpha}-\left(\frac{\mmin}{\Msun}\right)^{1-\alpha}\right]$. Later when $\left(\frac{\mmax}{\Msun}\right)^\gamma-\left(\frac{\mmin}{\Msun}\right)^\gamma>\frac{\gamma t}{t_0}\gg \left(\frac{\mmin}{\Msun}\right)^\gamma$, the total number of clusters is $N_{\rm imp}(t)\sim\frac{1}{1-\alpha}\left[\left(\frac{\mmax}{\Msun}\right)^{1-\alpha}-\left(\frac{\gamma t}{t_0}\right)^{(1-\alpha)/\gamma}\right]$ and decreases with time as $t^{(1-\alpha)/\gamma}$.
Finally, when $t>t_{\rm max}\equiv\frac{t_0}{\gamma}\left[\left(\frac{\mmax}{\Msun}\right)^\gamma-\left(\frac{\mmin}{\Msun}\right)^\gamma\right]$ (equivalent to $M_{\rm up}<\mmin$), all clusters evolve to a mass below $\mmin$ and therefore $N_{\rm imp}(t)=0$. 

\subsubsection{Continuous injection case}

\begin{figure}
\includegraphics[width=\columnwidth]{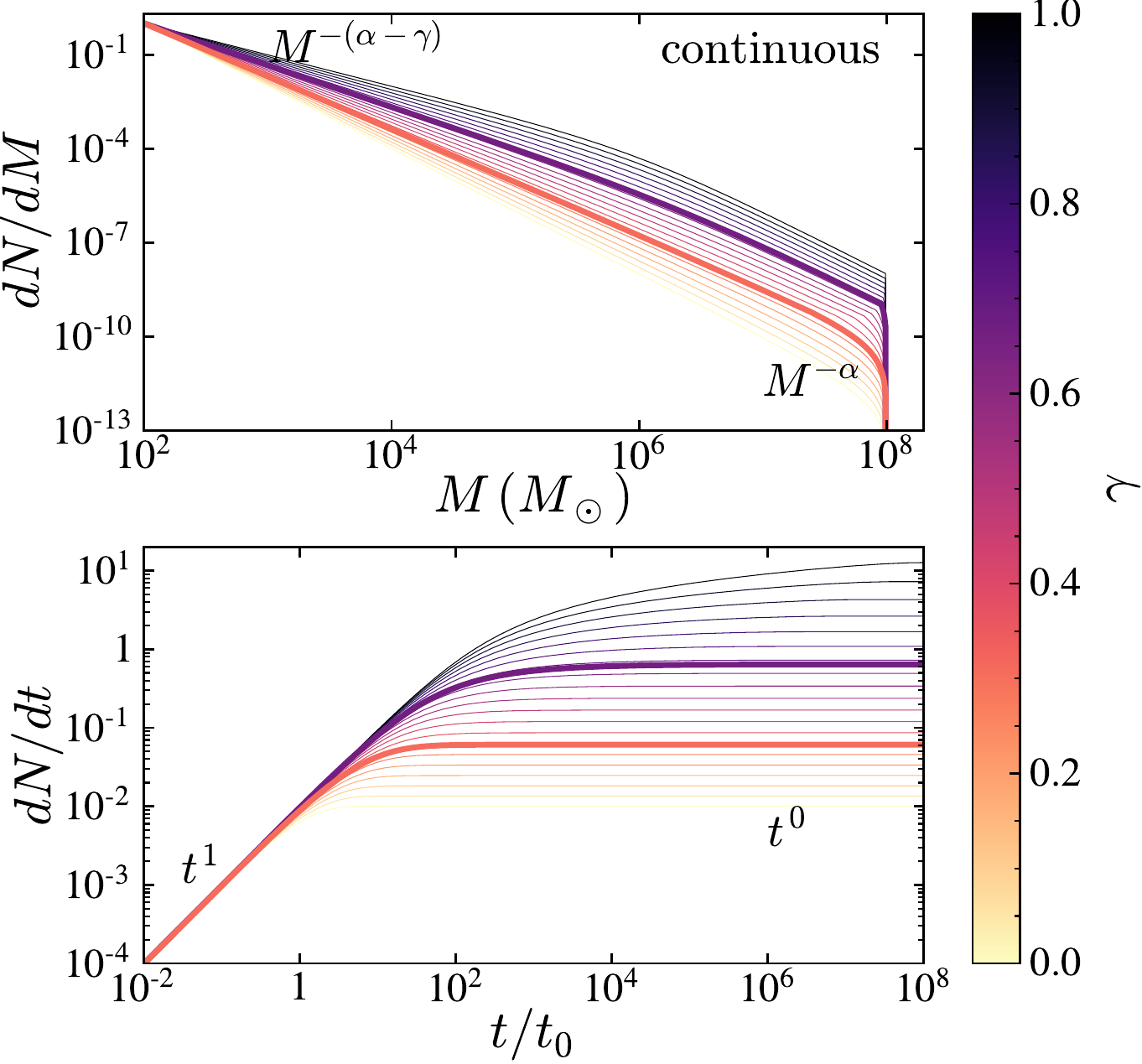}
\vspace{0mm}
\caption{The same as \autoref{fig:analytical-YMC-impulsive}, but for the continuous injection case.}
  \label{fig:analytical-YMC-continuous}
\end{figure}
Now we study the time evolution of the mass function and number of clusters with a \textit{constant} cluster formation rate over time by assuming all clusters follow the same CIMF ($dN/dM\propto M^{-\alpha}$). We call this the continuous injection case. The evolved mass function at time $T$ for the whole cluster population can be obtained by integrating \autoref{eq:dndmdt} the whole YMC age range from 0 to $T$:
\begin{equation}
    \frac{dN_{\rm cont}}{dM}=\int_0^{{\rm Min}[T,t_{\rm up}(M)]} \frac{dN_{\rm imp}(M,t)}{dMdt}dt,
\end{equation}
where $t_{\rm up}(M)=\frac{t_0}{\gamma}\left[\left(\frac{\mmax}{\Msun}\right)^\gamma-\left(\frac{M}{\Msun}\right)^\gamma\right]$ is the time when the mass of most massive clusters evolves from $\mmax$ to $M$. If $T>t_{\rm up}(M)$,
\begin{multline}
    \frac{dN_{\rm cont}}{dM}=\int_0^{t_{\rm up}} \frac{dN_{\rm imp}(M,t)}{dMdt}dt\propto \\ \frac{t_0}{1-\alpha}\left(\frac{M}{M_\odot}\right)^{\gamma-\alpha}\left[\left(\frac{\mmax}{\Msun}\right)^{1-\alpha}-1\right].
\end{multline}
On the other hand, if $T<t_{\rm up}(M)$,
\begin{multline}
    \frac{dN_{\rm cont}}{dM}=\int_0^T \frac{dN_{\rm imp}(M,t)}{dMdt}dt\propto \\ \frac{t_0}{1-\alpha}\left(\frac{M}{M_\odot}\right)^{\gamma-\alpha}
    \left\{\left[1+\frac{\gamma T}{t_0}\left(\frac{M}{M_\odot}\right)^{-\gamma}\right]^{\frac{1-\alpha}{\gamma}}-1\right\}
\end{multline}
In the high-mass end (or younger population) of the mass function where $\gamma T/t_0(M/M_\odot)^{-\gamma}\ll1$, the slope of the mass function is $\alpha$, the same as the CIMF. In the low-mass end (or older population), the slope of the mass function becomes $\alpha-\gamma$, and depends on both the power-law slope of the CIMF and the index of the mass-dependent tidal disruption timescale.

The time evolution of the number of clusters can be obtained by integrating \autoref{eq:nt-impulsive} over time:
\begin{multline}
    N(T)=\int_0^{{\rm Min}[T, t_{\rm max}]} N_{\rm imp}(t) dt \propto \\
    \frac{1}{1-\alpha}\left[\left(\frac{\mmax}{\Msun}\right)^{1-\alpha}t \,+ \right. \\ \left. -\frac{t_0}{1-\alpha+\gamma}\left(\left(\frac{\mmin}{\Msun}\right)^\gamma+\frac{\gamma t}{t_0}\right)^{(1-\alpha+\gamma)/\gamma}\right] \Biggr\rvert_0^{{\rm Min}[T, t_{\rm max}]}.
\end{multline}
For $T<t_{\rm max}$,
\begin{multline}
    N(T) =\int_0^{T} N_{\rm imp}(t) dt \propto \\
    \frac{1}{1-\alpha}\left\{ \mmax^{1-\alpha}T \, + \right. \\ \left. - \mmin^{1-\alpha+\gamma}\frac{t_0}{1-\alpha+\gamma}\left[(1+\frac{\gamma T}{t_0}\mmin^{-\gamma})^{(1-\alpha+\gamma)/\gamma}-1\right]\right\}.
\end{multline}
Therefore, at early times, when $t\ll\frac{t_0}{\gamma}(\frac{M_{\rm min}}{M_\odot})^{-\gamma}$, the number of clusters, $N(T)\propto\frac{T}{1-\alpha}(\mmax^{1-\alpha}-\mmin^{1-\alpha}) \propto T$, increases linearly with time.
On the other hand, for $T\geq t_{\rm max}$, the total number of clusters,
\begin{multline}
    N(T) =\int_0^{t_{\rm max}} N_{\rm imp}(t) dt \propto
    \frac{t_0}{\gamma(1-\alpha)(1-\alpha+\gamma)} \\
    \left[(1-\alpha)\mmax^{1-\alpha}(\mmax^{\gamma}-\mmin^{\gamma}) - \gamma\mmax^{\gamma}(\mmax^{1-\alpha}-\mmin^{1-\alpha}) \right],
\end{multline}
is a constant, suggesting the number of clusters reaches an equilibrium as the cluster injection rate equals the disruption rate.

\bsp	
\label{lastpage}
\end{document}